\documentclass[iicol, pdflatex,sn-mathphys-num]{sn-jnl}% Math and Physical Sciences 

\usepackage{graphicx}%
\usepackage{multirow}%
\usepackage{amsmath,amssymb,amsfonts}%
\usepackage{amsthm}%
\usepackage{mathrsfs}%
\usepackage[title]{appendix}%
\usepackage{xcolor}%
\usepackage{bm}
\begin{document}

\title[On the planar impact of a disc and a square particle considering the impact angle and particle orientation]{On the planar impact of a disc and a square particle considering the impact angle and particle orientation}

\author*[1]{\fnm{Dominik} \sur{Krengel}}\email{krengel.dominik@u.tsukuba.ac.jp}

\affil*[1]{\orgdiv{Department of Engineering Mechanics and Energy}, \orgname{University of Tsukuba}, \orgaddress{\street{1-1-1,
Tennodai}, \city{Tsukuba}, \postcode{305-8573}, \state{Ibaraki}, \country{Japan}}}

\abstract{Despite the simple impact of a rigid particle being a centuries old problem, a conclusive treatment of the general case is still outstanding. The influences of particle shape as well as elastic and plastic deformation of the particle upon contact significantly complicate the problem. Experiments have shown the possibilities of backward movement as well as forward movement with backward rotation after an impact. This paper investigates the planar impact of a disc and square particle under systematic consideration of the impact angle and initial particle orientation. We investigate the kinematic variables of the particle both post-impact as well as their evolution during contact. For impacts with friction, we find oscillatory behaviour of the tangential contact force in both discs and squares depending on the impact angle and particle orientation. We further demonstrate the possibility of backwards deflection for square particles with backward orientation under steep impact angles, which is not possible for round particles under the same conditions. In contrast, high deflection angles are achieved with a kind of “double-impact” where the centroid of the particle moves away from the plane so that the surface detaches first, but reconnects again due to induced rotation, which also is not possible for round particles. Lastly, we present different coefficients of restitution for the square particle with respect to the combined influence of impact angle and particle orientation upon impact.}


\maketitle

\section{Introduction}
The dynamics of a non-static granular aggregate is, at its core, governed by a series of dissipative impact events between two rigid bodies, whether one considers extensive multibody systems with thousands or even millions of particles and collisions\,\cite{Jiang2025,Chen2024}, or only follows the kinematics of a single particle\,\cite{Fernandez2021,Krengel2024,Lyu2025}. Although the planar impact of a rigid body has been studied for centuries (see\,\cite{Goldsmith1999} and references therein) the topic is far from being closed. Work continues primarily across three different fields, which are, lamentably, largely ignorant of each other due to their different premises and focus.

In statistical mechanics, the pre- and post collision kinematics of a particle impact (particle-particle and particle-plane) are usually quantified by means of normal and tangential restitution coefficients\,\cite{Wu2003,Weir2005}. A variety of different formulations for both coefficients have been derived for simple spheres\,\cite{Schwager2007,Becker2008,Mueller2011}, but also for more elaborate shapes\,\cite{Ji2020,Wedel2024}. Statistical analysis of these models have shown a non-trivial relation between the rectilinear and angular velocities\,\cite{Hastie2013,Glielmo2014,Zhang2020,Fernandez2021}, with some rather surprising results\,\cite{Saitoh2010,Mueller2012}. While not strictly true\,\cite{Schaefer1996,Montaine2011}, the coefficients of restitution are generally used as an invariant material parameter to simplify the analysis of contact processes in e.g. kinetic theory or event-driven simulations. 

Rockfall modelling, as a subset of civil engineering, often uses restitution coefficients\,\cite{Chau2002} together with simplified particle shape, motion and impact laws to describe downslope rock motion\,\cite{Duan2023}. However, rocks are macroscopic objects with complex, non-spherical shapes that complicate the influence of impact angle and initial angular velocity on the kinematics\,\cite{Asteriou2018,Wang2018,Leine2021,Ge2024}. Experimental and numerical modelling observed for example backward motion of the block upon impact\,\cite{Krengel2024}, or forward rectilinear motion combined with backward rotation\,\cite{Ji2023}. The predictive power of modelling is therefore limited, and work is ongoing to develop more elaborate formulations of the coefficient of restitution\,\cite{Dattola2021,Dattola2023} for a better prediction of the post-collision rebound kinematics of rocks.

In the field of impact mechanics, the planar impact of round (or, occasionally, arbitrary shaped) particles has been studied mathematically\,\cite{Stronge2000,Nordmark2009,Glocker2012,Glocker2013,Lyashenko2015}. While the contact can, in principle, be treated with a continuum approach\,\cite{Shen2011}, the problem is usually reduced to the kinematics of a single contact point\,\cite{Brach1989,Jia2017} where the surrounding deformation is neglected.
 For the majority of impact mechanics approaches, the coefficient of restitution is again considered as independent of the impact conditions\,\cite{Stronge1994,Stronge2001}. Impacts are classified into several groups, whose correct identification of the contact mode is essential for their study\,\cite{Wang1992}. For specific impact conditions, even reversal of the sliding velocity is possible\,\cite{Pfeiffer2017}. Despite the simple statement of the problem, the evolution of the contact point is far from simple, and various combinations of sliding, sticking or velocity reversal may occur during the impact\,\cite{Brach1993}, significantly altering the post-collision kinematics.

Although slowly changing, common to all three fields is that the shape of the impacting particle, as well as the consequences of shape, is largely omitted from consideration. While for spherical particles the rotational symmetry significantly simplifies analytical description and numerical computation, the actual shape of realistic particles may be rather complex and preclude analytical treatment. On the other hand, discrete element methods with polytopes allow representation of realistic particle shapes and thus the study of more complex impact conditions.

In this paper we investigate the planar impact of a disc and a square in two dimensions by means of a two-dimensional discrete element method. We focus both on the post-impact kinematics as well as the evolution of quantities during the course of the impact itself. The results are presented with respect to the impact angle for a disc and in terms of the particle orientation upon contact and impact angle for a square.

\section{DEM-simulation}
We employ a hard-particle soft-contact scheme to study the planar impact of squares and disc-shaped particles, based on the monograph by Matuttis and Chen\,\cite{Matuttis2014}.

\subsection{Methodology}
\begin{figure}[h]
 \centering
 \includegraphics[width=\columnwidth]{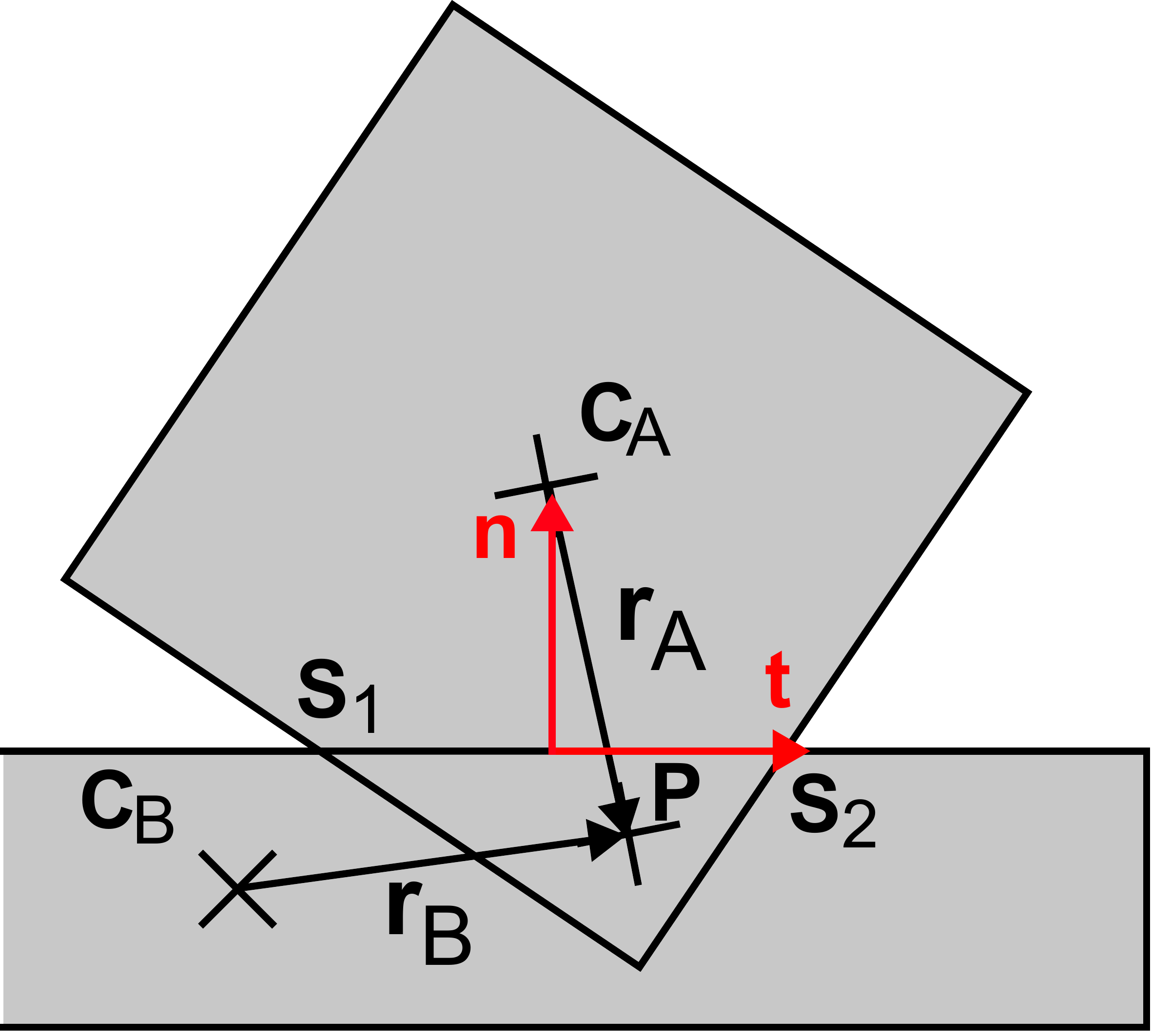}
 \caption{Contact geometry for two polygonal particles with the centres of mass $\mathbf{C}_{\mathrm{A,B}}$, the overlap centroid $\mathbf{P}$, and the contact vectors $\mathbf{r}_{\mathrm{A,B}}$. Shown in red are the normal $\mathbf{n}$ and tangential $\mathbf{t}$ direction of the contact, as defined by the intersection points $\mathbf{S}_{1,2}$. The overlap is significantly exaggerated for the purpose of illustration.}
 \label{fig:contactgeometry}
\end{figure}

We calculate the interaction force between two particles as a ``penalty'' based on the particle overlap, as a measure of the physical deformation of the particles, see Fig.\,\ref{fig:contactgeometry}. The force point is given by the centroid $\mathbf{P}$ of the overlap area. The connecting line between the intersection points of the two polygons, $\mathbf{S}_{1}$ and $\mathbf{S}_{2}$, defines the tangential direction $\mathbf{t}$, so the normal direction $\mathbf{n}$ is also fixed. The elastic force
\begin{equation}
 F_{\mathrm{el}} = \frac{YA}{l} 
\end{equation}
is proportional to the overlap area $A$ and the Young's modulus $Y$ and acts in normal direction. The characteristic length
\begin{equation}
 l = 4\frac{|\mathbf{r}_{\mathrm{A}}||\mathbf{r}_{\mathrm{A}}|}{|\mathbf{r}_{\mathrm{A}}|+|\mathbf{r}_{\mathrm{A}}|}
\end{equation}
for the contact vectors $\mathbf{r}_{\mathrm{A,B}}$ between the centres of mass $\mathbf{C}_{\mathrm{A,B}}$ to $\mathbf{P}$ allows to define the force in units of [N]. Additionally, a dissipative force is acting in normal direction,
\begin{equation}
 F_{\mathrm{diss}} = \gamma\sqrt{mY}\frac{\dot{A}}{l},
\end{equation}
with the reduced mass
\begin{equation}
 \frac{1}{m} = \frac{1}{m_{\mathrm{A}}} + \frac{1}{m_{\mathrm{B}}},
\end{equation}
and a damping constant $\gamma$. The total normal force is then
\begin{equation}
 F_{\mathrm{N}} = F_{\mathrm{el}} + F_{\mathrm{diss}}.
\end{equation}
As the dissipative force may not overcompensate the elastic force, a cutoff is required, see\,\cite{Matuttis2014} for details. Friction, acting in tangential direction, is given by the Cundall-Strack model\,\cite{Cundall1979},
\begin{equation}
 F_{\mathrm{T}}(t) = \left\{\begin{array}{ll}
        F_{\mathrm{T}}(t-\mathrm{d}t)-k_{\mathrm{t}}v_{\mathrm{t}}\mathrm{d}t & |F_{\mathrm{T}}(t)|\leq \mu F_{\mathrm{N}}\\
				\mathrm{sgn}\left(F_{\mathrm{T}}(t-\mathrm{d}t)\right)\mu F_{\mathrm{N}} & |F_{\mathrm{T}}(t)| > \mu F_{\mathrm{N}}
\end{array}\right.,
\end{equation}
where $v_{\mathrm{t}}$ is the relative tangential velocity at the contact point and $k_{\mathrm{t}}$ the ``tangential stiffness'' of a spring. Lastly, the torques are computed based on the total force $\mathbf{F} = F_{\mathrm{N}}\mathbf{n} + F_{\mathrm{T}}\mathbf{t}$ and the contact vectors as
\begin{equation}
 \bm{\tau}_{\mathrm{A,B}} = \mathbf{r}_{\mathrm{A,B}}\times\mathbf{F}.
\end{equation}

\subsection{Setup}
We consider two particle shapes: firstly a disc, which is modelled as a regular convex polygon with 120 corners, and secondly a regular convex square. Both particles are created with the same circum-radius. The particles are placed a fixed distance from the plane with an initial velocity
\begin{equation}
 \left(\begin{array}{c}v_{x}(0)\\ v_{y}(0)\end{array}\right) = v(0)\left(\begin{array}{c}\cos{\theta_{\mathrm{in}}}\\ \sin{\theta_{\mathrm{out}}}\end{array}\right),
\end{equation}
where $v(0)=0.05$ [m/s] is the initial velocity magnitude, so that for all cases, the time until impact remains the same. We primarily focus on the impact of frictional particles with contact damping in normal direction. Additionally we consider the impact of dissipative particles without rotational degrees of freedom and non-dissipative particles to contrast our results.

\begin{figure}[h]
 \centering
 \includegraphics[width=\columnwidth]{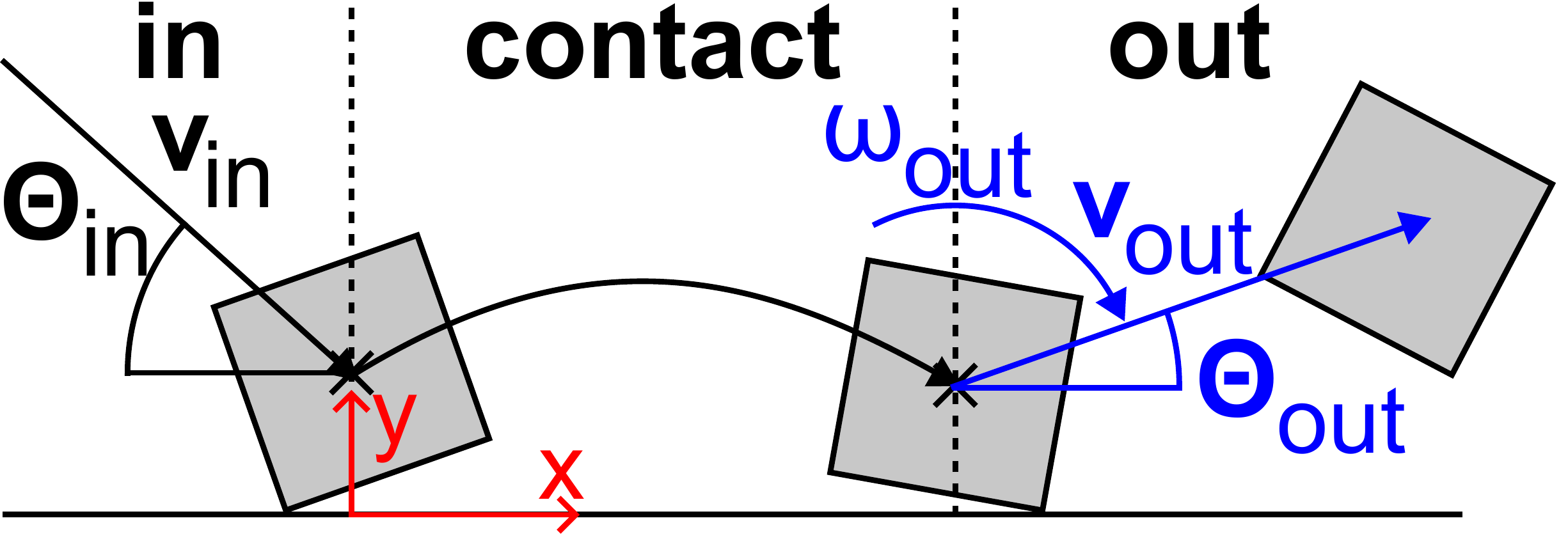}
 \caption{Reference system for the block impact, showing the initial (black) and rebound (blue) variables of the particle. For a planar impact the normal direction is aligned with the y-direction, and the tangential direction is aligned with the x-direction. Impact and rebound angle are measured with respect to the impact plane. Rebound angles larger than 90$^\circ$ indicate backwards bouncing.}
 \label{fig:impactgeometry}
\end{figure}

Figure\,\ref{fig:impactgeometry} shows the geometry of the setup. Impact angle $\theta_{\mathrm{in}}$ and rebound angle $\theta_{\mathrm{out}}$ are measured from the plane to the normal direction. Rebound angles larger than 90$^{\circ}$ indicate backwards deflection of the particle. The impact plane is aligned with the horizontal direction, so the normal and tangential contact directions are identical to the y- and x- direction respectively.

\begin{figure}[h]
 \centering
 \includegraphics[width=\columnwidth]{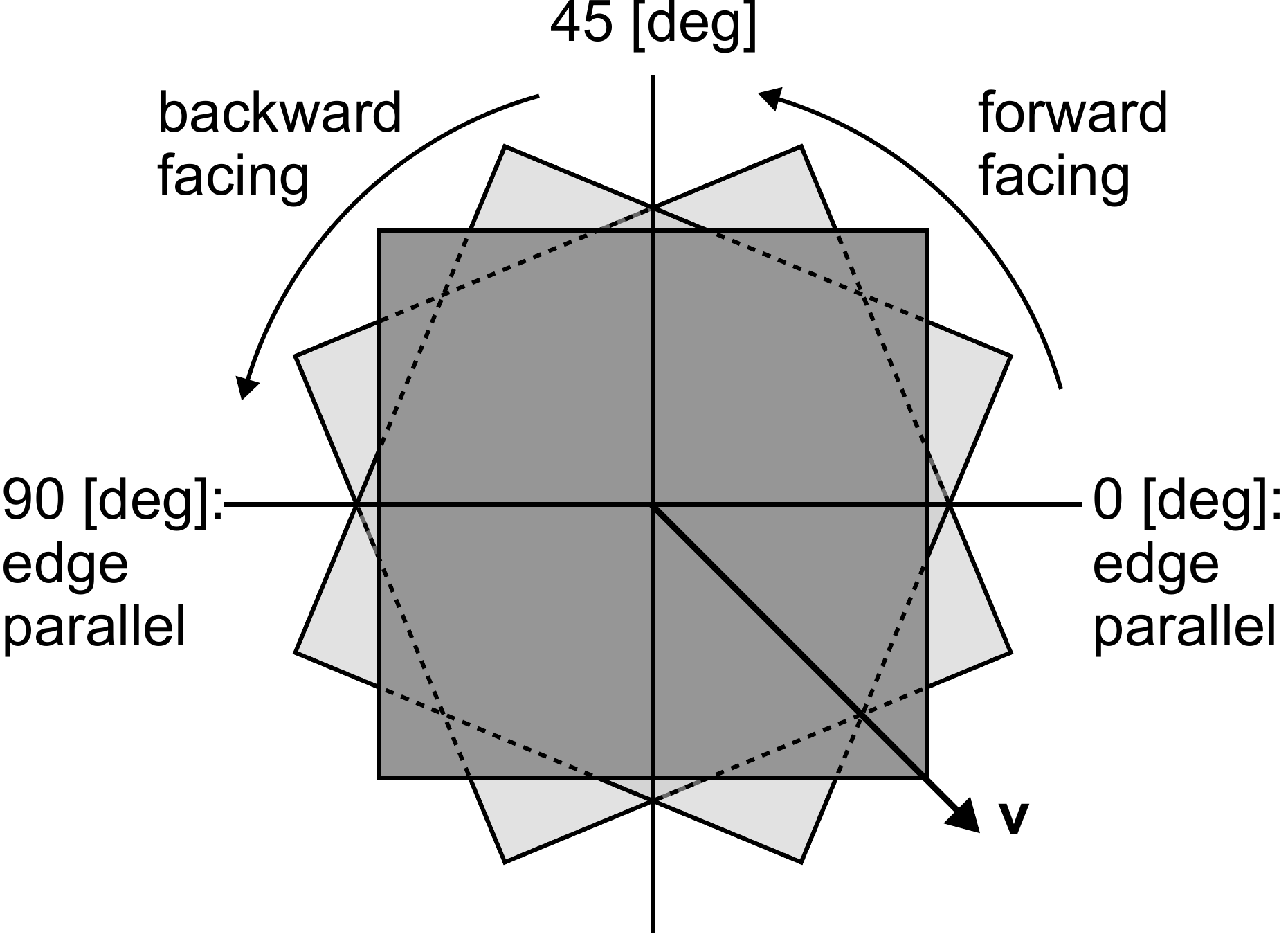}
 \caption{Block orientation $\alpha$ for the square particle, measured with respect to the angle between particle edge and the impact plane. The particle is considered forward facing when the second lowest edge is first in the direction of travel and backwards facing when it is trailing behind. Likewise, forward rotation is understood as turning towards the direction of travel, i.e. as negative values of $\omega$.}
 \label{fig:square_orientation}
\end{figure}

During the study we systematically vary the impact angle $\theta_{\mathrm{in}}$ between 17$^\circ$ and 89$^\circ$. Angles below 17$^\circ$ are omitted as otherwise the particle will be initialized already in contact with the plane, given the same initial constraints as for other angles. For square particles, we further systematically change the particle orientation within their symmetry angle, see Fig.\,\ref{fig:square_orientation}
During the simulation gravity is set to zero to omit its influence on the impact process. The particles are initialized without angular velocity, but may start to rotate upon contact if the specific test case allows it. If not mentioned otherwise, the parameters in table\,\ref{tab:parameters} are used.


\begin{table}[h]
\centering
\begin{tabular}{l|c|c}
 Timestep & d$t$ & $5\cdot10^{-6}$ [s]\\
 Gravity  & $g$  & 0\\
 Friction & $\mu$ & 0.5\\
 Damping  & $\gamma$ & 0.5\\
 Young's modulus & $Y$ & $10^{9}$ [N/m]\\
 Density & $\varrho$ & 5000 [kg/m$^2$]\\
 Initial velocity & $v(0)$ & 0.05 [m/s]\\
 Initial angular velocity & $\omega(0)$ & 0 [1/s]\\
 Impact angle & $\theta_{\mathrm{in}}$ &17-89$^{\circ}$
\end{tabular}
\caption{Simulation parameters used in this study.}
\label{tab:parameters}
\end{table}

\section{Impact of a disc}
The most simple impact model is that of a point particle colliding with a flat plane. Particle rotation and deformation upon impact are not considered, and so the rebound angle will be equal to the impact angle. If the collision is dissipative, then the post-collision velocity will be less than the pre-collision velocity, otherwise it will be equal. 

For a disc (or sphere in three dimensions), elastic deformation is possible, friction will additionally dissipate energy, and the particle can roll along the plane, all of which complicated the impact behaviour and post-collision properties of the particle.

\subsection{Post-collision behaviour}
\begin{figure}[t]
 \centering
 \includegraphics[width=\columnwidth]{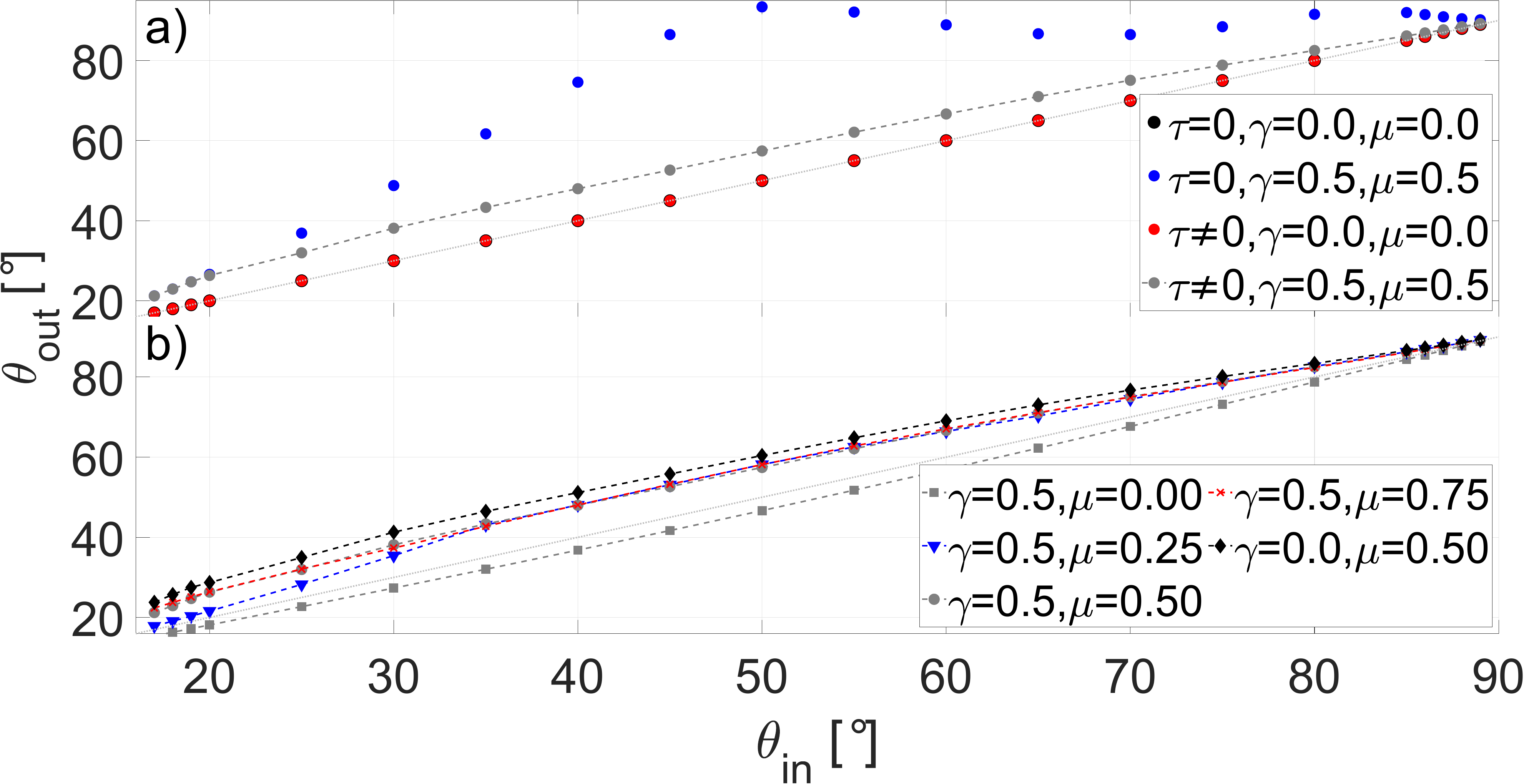}%
 \caption{a) Rebound angle $\theta_{\mathrm{out}}$ as a function of the impact angle $\theta_{\mathrm{in}}$ for different test cases. The dashed line indicates $\theta_{\mathrm{out}} = \theta_{\mathrm{out}}$. Backwards deflection was observed only for dissipative systems without rotational degrees of freedom (indicated as $\tau=0$). b) Variation of $\theta_{\mathrm{out}}$ for different coefficients of friction $\mu$, and for friction without normal damping ($\gamma = 0.0, \mu=0.5$)}
 \label{fig:N120_reboundangle}
\end{figure}

In the absence of any dissipative forces ($\mu=0,\gamma=0$), the rebound angle $\theta_{\mathrm{out}}$ is equivalent to the impact angle $\theta_{\mathrm{in}}$, as shown in Fig.\,\ref{fig:N120_reboundangle}, a). For particles with dissipative forces and rotation ($\mu=0.5,\gamma=0.5,\tau\neq0$) we find that $\theta_{\mathrm{out}} \gtrsim \theta_{\mathrm{in}}$, but converges against $\theta_{\mathrm{in}}$ at very low or very high impact angles. In addition, for particles with dissipation, but with disabled rotational degrees of freedom ($\tau=0,\mu=0.5,\gamma=0.5$), $\theta_{\mathrm{out}}$ rapidly increases until around $\theta_{\mathrm{in}}\sim50^{\circ}$ where the particle will begin deflecting backwards and oscillate around $\theta_{\mathrm{out}}=90^{\circ}$. Finally, as we have modelled our discs as polygons with 120 corners we note that the initial orientation of the particles does not affect the results of the simulations. The standard deviation across multiple different initial orientations ranges between 0.05 and 0.29.

To clarify the influence of friction, we further vary the friction coefficient between $\mu=0$ and $\mu=0.75$, see Fig.\,\ref{fig:N120_reboundangle}, b). At large impact angles $\theta_{\mathrm{out}}$ for different values of $\mu>0$ follows the same curve. However, at lower values of $\theta_{\mathrm{in}}$, $\theta_{\mathrm{out}}$ starts to diverge from a shared form. The divergence occurs at higher values of $\theta_{\mathrm{in}}$ for $\mu\rightarrow 0$ and at lower values for $\mu\rightarrow 1$. For frictionless particles with normal damping (i.e. $\mu=0.0, \gamma=0.5$) we find that $\theta_{\mathrm{out}} < \theta_{\mathrm{in}}$, but converges to $\theta_{\mathrm{out}}=\theta_{\mathrm{in}}$ at steep or flat impact angles. Lastly, in the case of frictional, but undamped particles ($\mu=0.5,\gamma=0.0$) tends to deviate more from $\theta_{\mathrm{out}}=\theta_{\mathrm{in}}$ as for the same $\mu$ with finite $\gamma$.

 \begin{figure}[t]
 \centering
 \includegraphics[width=\columnwidth]{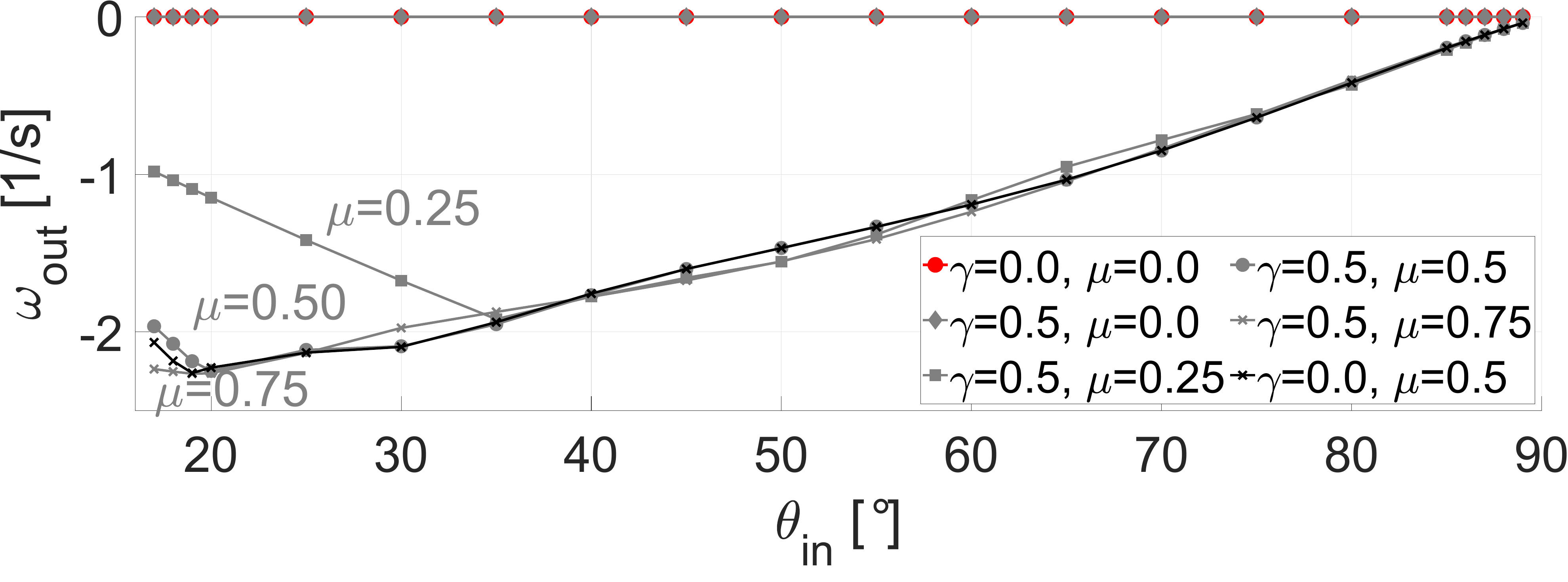}%
 \caption{Post-impact angular velocity $\omega_{\mathrm{out}}$ as a as a function of the impact angle $\theta_{\mathrm{in}}$ for the test cases with rotational degrees of freedom. Impacts with higher ($\mu = 0.75$) and lower ($\mu=0.25$) friction coefficients collapse onto the same curve for steeper impact angles $\theta_{\mathrm{in}}$, but separate at lower impact angles.}
 \label{fig:N120_omega}
\end{figure}
\begin{figure}[b]
 \centering
 \includegraphics[width=\columnwidth]{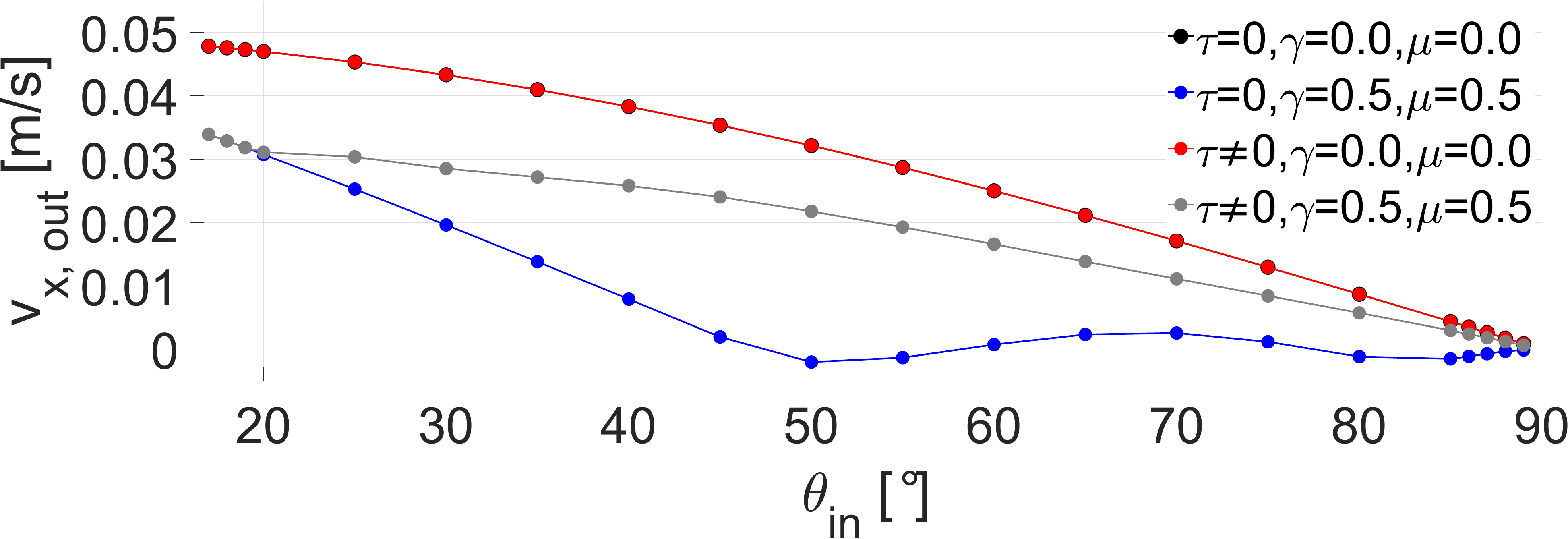}%
 \caption{Post-impact translational velocity component $v_{x,\mathrm{out}}$ as a function of the impact angle $\theta_{\mathrm{in}}$. Cases without dissipation follow the same curve. Cases with dissipation follow different curves in their tangential components whether rotation is allowed or not.}
 \label{fig:N120_vx}
\end{figure}
As a tangential force is required to induce rotation, frictionless particles ($\mu=0$) do not obtain any angular velocity $\omega$ after the impact (Fig.\,\ref{fig:N120_omega}). For systems with friction the particle always rotates in forward direction ($\omega<0$) after the impact, regardless of the impact orientation, while the rotation magnitude $|\omega|$ increases up to a $\mu$-dependent maximum. The decrease of $|\omega|$ for low and high impact angles is in agreement with experiments performed by Gorham and Kharaz\,\cite{Gorham2000} and analytical solutions obtained by Lyashenko et al.\,\ref{Lyashenko2017}. The peak angular velocity magnitude is found at larger $\theta_{\mathrm{in}}$ if $\mu$ decreases and conversely at lower $\theta_{\mathrm{in}}$ if $\mu$ increases. Notably, the data for $|\omega_{\mathrm{out}}|$ beyond its maximum value collapse onto the same curve for all values of $\mu>0$, regardless of the presence or absence of contact damping $\gamma$.

Without dissipative forces, the tangential velocity at separation, $v_{x,\mathrm{out}}$, falls onto the same curve, regardless if particle rotation is possible ($\tau\neq 0$) or not ($\tau=0$), see Fig.\,\ref{fig:N120_vx}.
If the particle is dissipative, then the tangential velocity at separation is the same for systems with and without rotation at very flat impacts ($\theta_{\mathrm{in}}\lesssim20^{\circ}$), but separates for larger $\theta_{\mathrm{in}}$. $v_{x,\mathrm{out}}$ quickly decreases to zero at around $\theta_{\mathrm{in}}\sim50^{\circ}$ if rotation is not possible and oscillates around zero as $\theta_{\mathrm{in}}$ increases further, mirroring the behaviour of the rebound angle $\theta_{\mathrm{out}}$ shown in Fig.\,\ref{fig:N120_reboundangle}, a). Changes in $\mu$ lead to the same trends in $v_{x,\mathrm{out}}$ as in $\omega_{\mathrm{out}}$: At $\mu$-dependent values of $\theta_{\mathrm{in}}$ $v_{x,\mathrm{out}}$ branches out from a common curve at larger $\theta_{\mathrm{in}}$. Lastly, we found that the normal rebound velocity only depends on the presence or absence of contact damping $\gamma$.

\begin{figure}[t]
 \centering
 \includegraphics[width=\columnwidth]{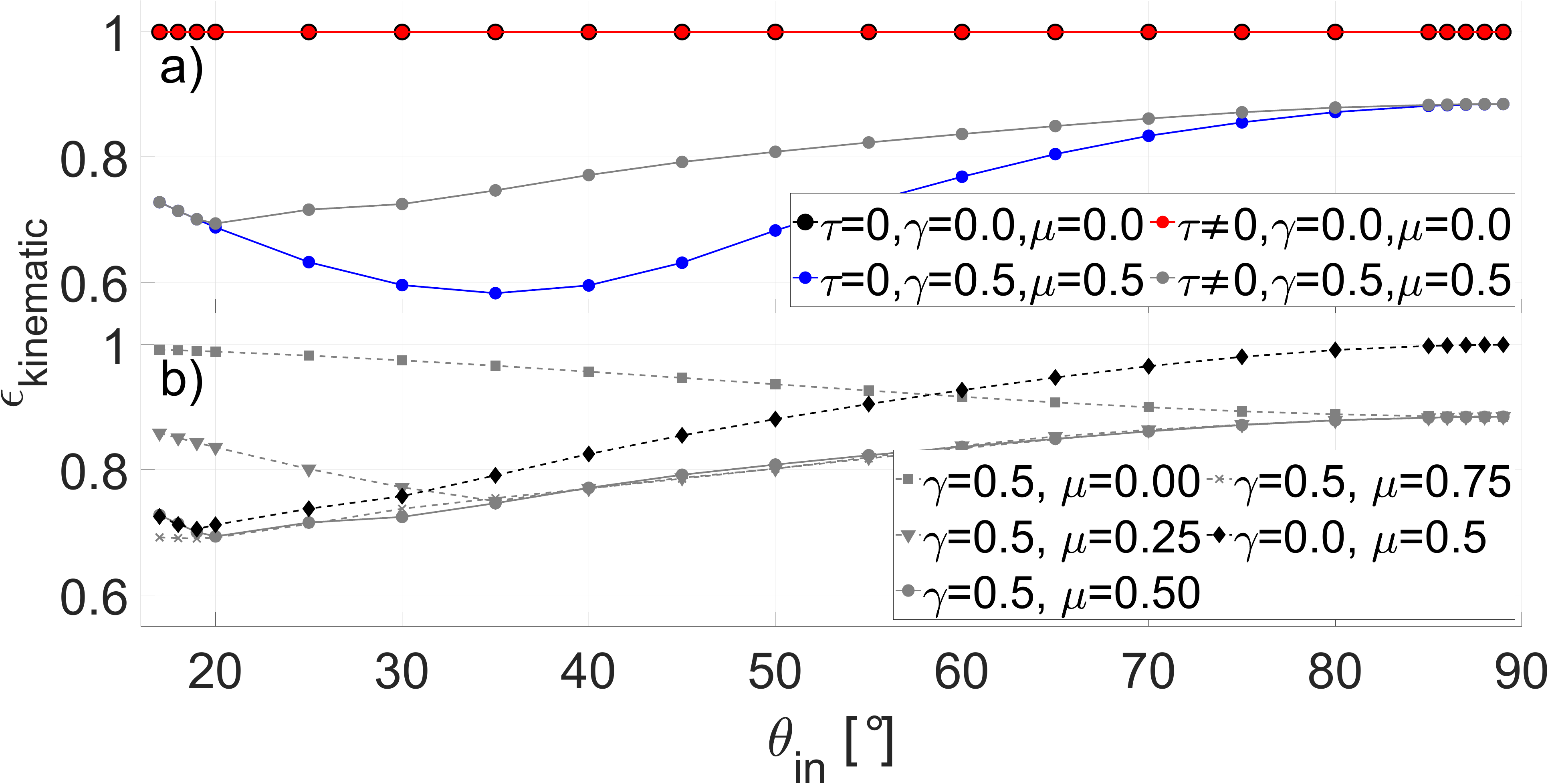}%
 \caption{Kinematic $\varepsilon_{\mathrm{kinematic}}$ restitution coefficient as function of the impact angle $\theta_{\mathrm{in}}$ for a) different conditions and b) different friction coefficients.}
 \label{fig:N120_restitution_kinematic}
\end{figure}Post-impact kinematics are commonly computed by a variety of restitution coefficients (e.g.\,\cite{Asteriou2015,Asteriou2018}). The kinematic coefficient of restitution,
\begin{equation}
 \varepsilon_{\mathrm{kinematic}} = \frac{v_{\mathrm{out}}}{v_{\mathrm{in}}}
 \label{eq:COR_kinematic}
\end{equation}
describes the momentum change of the particle during the impact.

Figure\,\ref{fig:N120_restitution_kinematic},a) shows that without any dissipative forces, $\varepsilon_{\mathrm{kinematic}}$ is conserved through the impact at any impact angle. For impacts with dissipation and rotational degrees of freedom, $\varepsilon_{\mathrm{kinematic}}$ decreases until around $\theta_{\mathrm{in}}\sim20^{\circ}$, then increases again before saturating at $\theta_{\mathrm{in}}=90^{\circ}$, which also agrees with the results obtained by Sch\"afer et. al. for the oblique impact of two discs\,\cite{Schaefer1996}. For impacts with dissipation but without rotational degrees of freedom, $\varepsilon_{\mathrm{kinematic}}$ follows the form of impacts with rotation, but decreases even further until a minimum at around $\theta_{\mathrm{in}}\sim35^{\circ}$. $\varepsilon_{\mathrm{kinematic}}$ without rotation then saturates at the same value as with rotation. If the coefficient of friction decreases, then the local minimum in $\varepsilon_{\mathrm{kinematic}}$ occurs at larger values of $\theta_{\mathrm{in}}$, if it increases, then the minimum shifts to smaller values, see Fig.\,\ref{fig:N120_restitution_kinematic}, b). If the particle is frictionless, then $\lim{\varepsilon_{\mathrm{kinematic}}}_{\theta_{\mathrm{in}}\rightarrow 0^{\circ}} = 1$, as the initial translational momentum is largely stored in the tangential velocity and thus largely unaffected by dissipation in normal direction. Likewise $\lim{\varepsilon_{\mathrm{kinematic}}}_{\theta_{\mathrm{in}}\rightarrow 90^{\circ}} < 1$ as here the majority of the 
initial translational momentum is stored in the normal velocity, and thus greatly affected by dissipation in normal direction. Correspondingly, in the absence of normal dissipation ($\gamma=0,\mu\neq0$) $\varepsilon_{\mathrm{kinematic}}$ follows the equivalent dissipative system but increases faster after the local minimum and saturates at 1 for $\theta_{\mathrm{in}} = 90^{\circ}$. Other coefficients of restitution are discussed in the appendix.

\subsection{Behaviour during contact}
During the collision, the disc experiences changing tangential (frictional) and normal forces. 
\begin{figure}[b]
\includegraphics[width=\columnwidth]{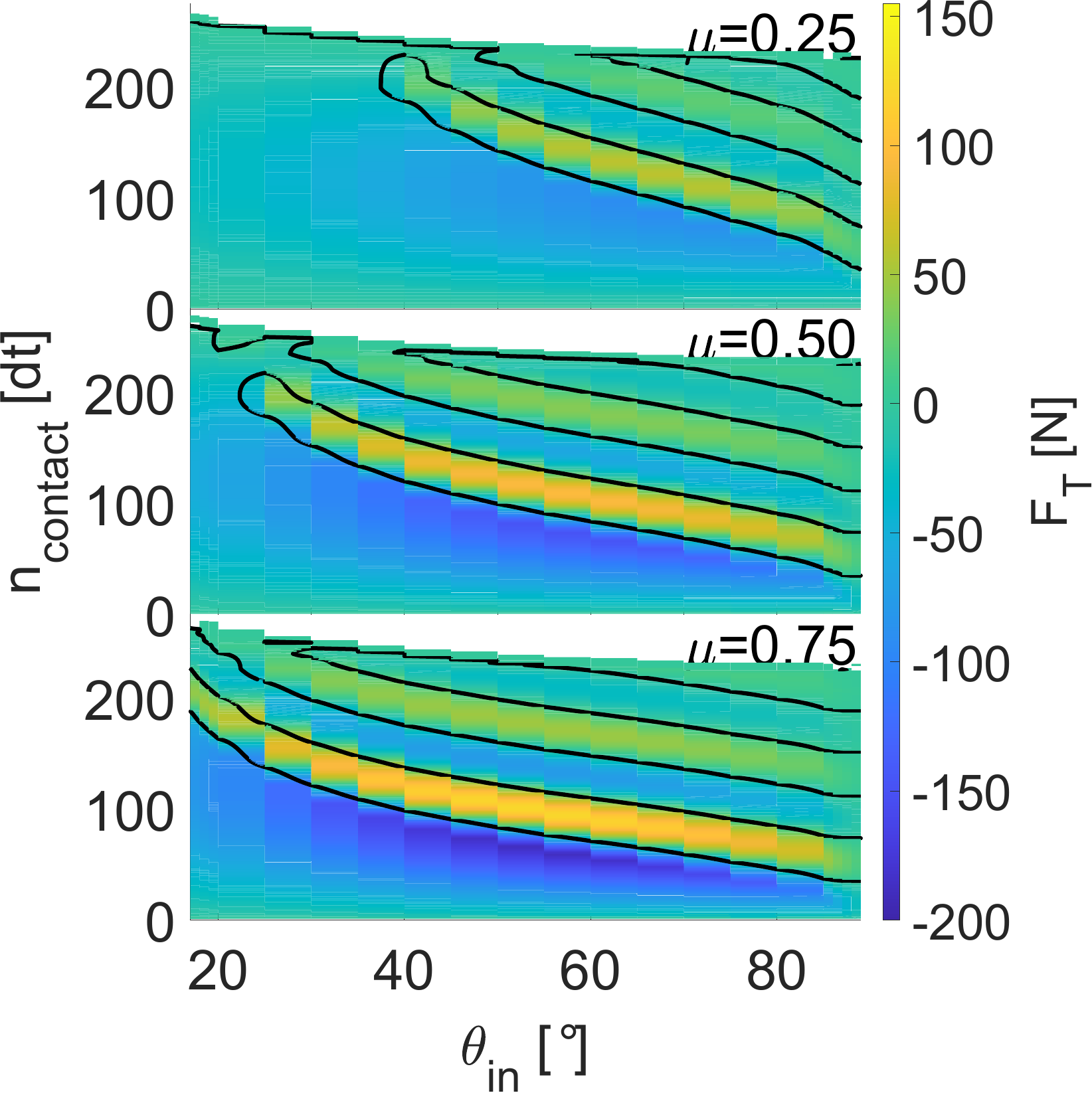}
\caption{Evolution of the tangential force $F_{T}$ acting on the disc during impact with the contact step $n_\mathrm{t}$ for different impact angles $\theta_{\mathrm{in}}$ and friction coefficients $\mu$. The black line marks the zero-transition of the force.}
\label{fig:N120_contact_FT}
\end{figure} 

In Fig.\,\ref{fig:N120_contact_FT} we plot the evolution of the induced tangential force for different impact angles $\theta_{\mathrm{in}}$ and friction coefficients. After an initial, strong deceleration ($F_{T}<0$), $F_{T}$ begins to oscillate around 0 in a decaying pattern of acceleration and deceleration. These oscillations appear as diagonal bands in the $\theta_{\mathrm{in}}-n_{\mathrm{contact}}$ space. Both the strength and and the $\theta_{\mathrm{in}}$-range for which these oscillations occur increase with larger $\mu$. If $\mu$ is smaller, then the oscillations occur only for steeper impacts, while for flatter impacts the tangential response remains negative. At the same time the acceleration increases towards the centre of each band. We found a similar force oscillation for frictional particles without rotation, see Fig.\,\ref{fig:N120_contact_FT_norot} in the appendix. For non-frictional particles the tangential force remained constant. 
\begin{figure}[t]
\includegraphics[width=\columnwidth]{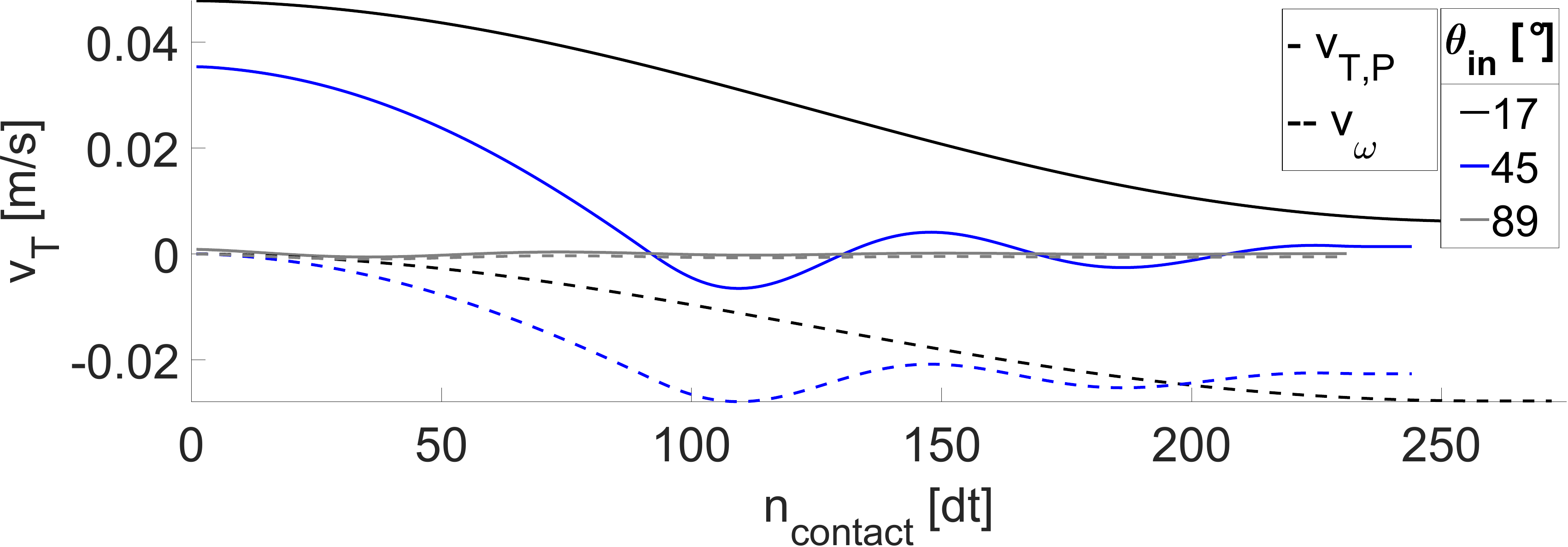}
\caption{Tangential velocity of the contact point (overlap centroid P) for flat impacts ($\theta_{\mathrm{in}} = 17^{\circ}$), steep impacts ($\theta_{\mathrm{in}} = 89^{\circ}$), and inbetween ($\theta_{\mathrm{in}} = 45^{\circ}$) for $\mu=0.5,\gamma=0.5$.}
\label{fig:N120_contactP_vt}
\end{figure}

We further observe a repeating reversal of the sliding velocity,
\begin{equation}
 v_{T,\mathrm{P}} = v_{x} -r_{\mathrm{A},y}\omega,
\end{equation}
in the same $\theta_{\mathrm{in}}$-range where we found the force oscillations, see Fig.\,\ref{fig:N120_contactP_vt}. The sliding velocity is oscillating around zero with decreasing amplitude until the contact terminates. Similar force oscillations and velocity reversal was found through dynamic analysis and experiment by Maw, Barber and Fawcett for the oblique impact of an elastic sphere sphere on a half space, which they attributed to tangential compliance\,\cite{Maw1976, Maw1981}. The same oscillations are found in the tangential velocity contribution of the angular velocity, $v_{\omega} = -r_{\mathrm{A},y}\omega$, although there the oscillations do not cross zero and stay strictly negative. For flat impacts, i.e. $\theta_{\mathrm{in}}$ less than required for the onset of oscillation, $v_{T,\mathrm{P}}$ monotonically declines towards a constant positive value before the contact terminates. The maximum amplitude of these oscillations decreases with increasing $\theta_{\mathrm{in}}$ as the initial tangential velocity decreases proportional to $\theta_{\mathrm{in}}$.

\begin{figure}[t]
\includegraphics[width=\columnwidth]{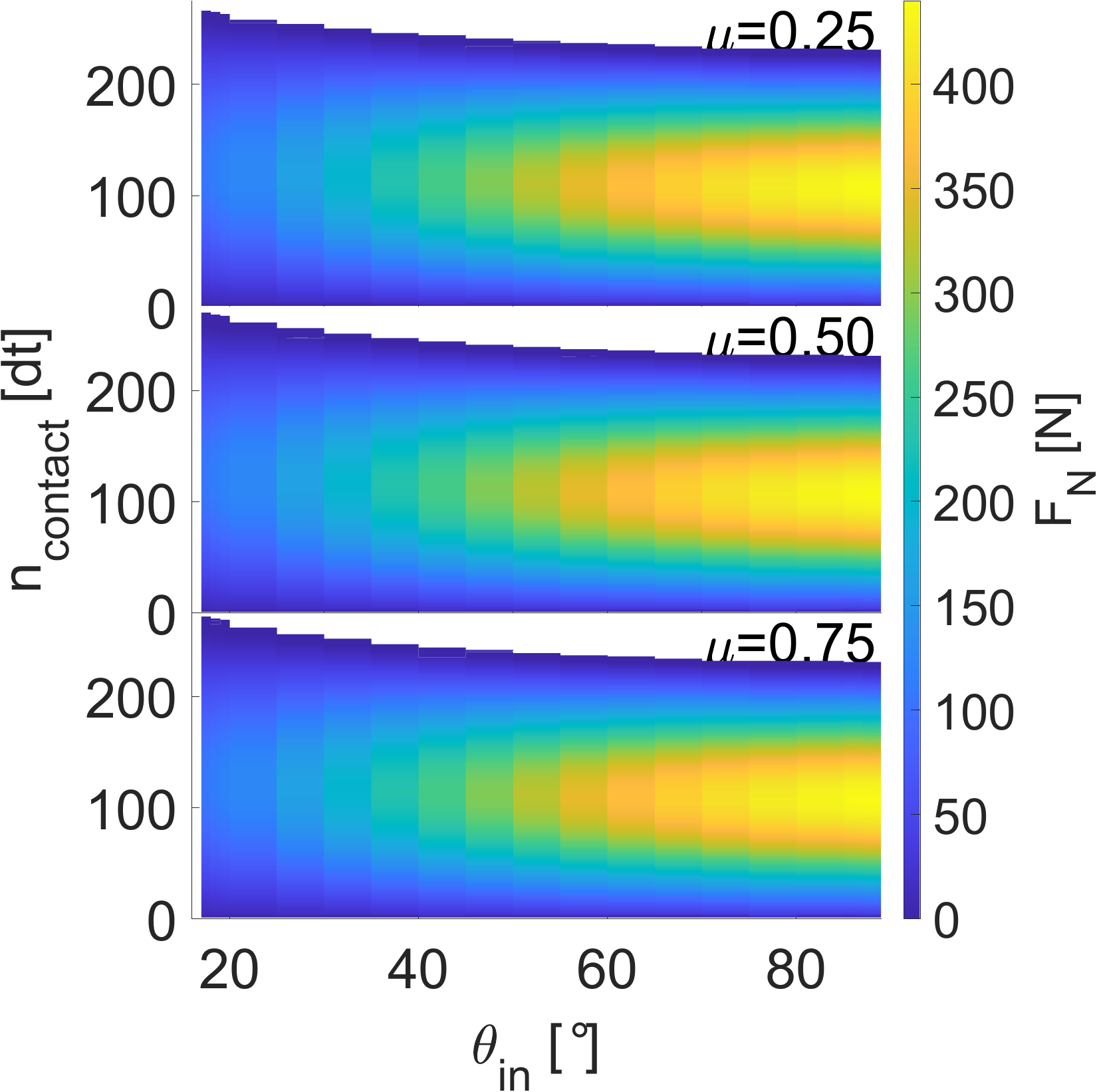}
\caption{Evolution of the normal force $F_{N}$ acting on the disc during impact with the contact step $n_\mathrm{t}$ for different impact angles $\theta_{\mathrm{in}}$ and friction coefficients $\mu$.}
\label{fig:N120_contact_FN}
\end{figure}

The corresponding normal force $F_{N}$ increases upon contact and decreases once the particle starts reversing (Fig.\,\ref{fig:N120_contact_FN}), corresponding to elastic compression /deformation and restitution respectively. For steeper impacts a greater response in $F_{T}$ is induced as more energy is carried in the normal component of the velocity, which leads to a greater deformation. The magnitude of the friction coefficient, the magnitude of the damping parameter, or the ability to rotate do not affect the normal force.

\begin{figure}[b]
\includegraphics[width=\columnwidth]{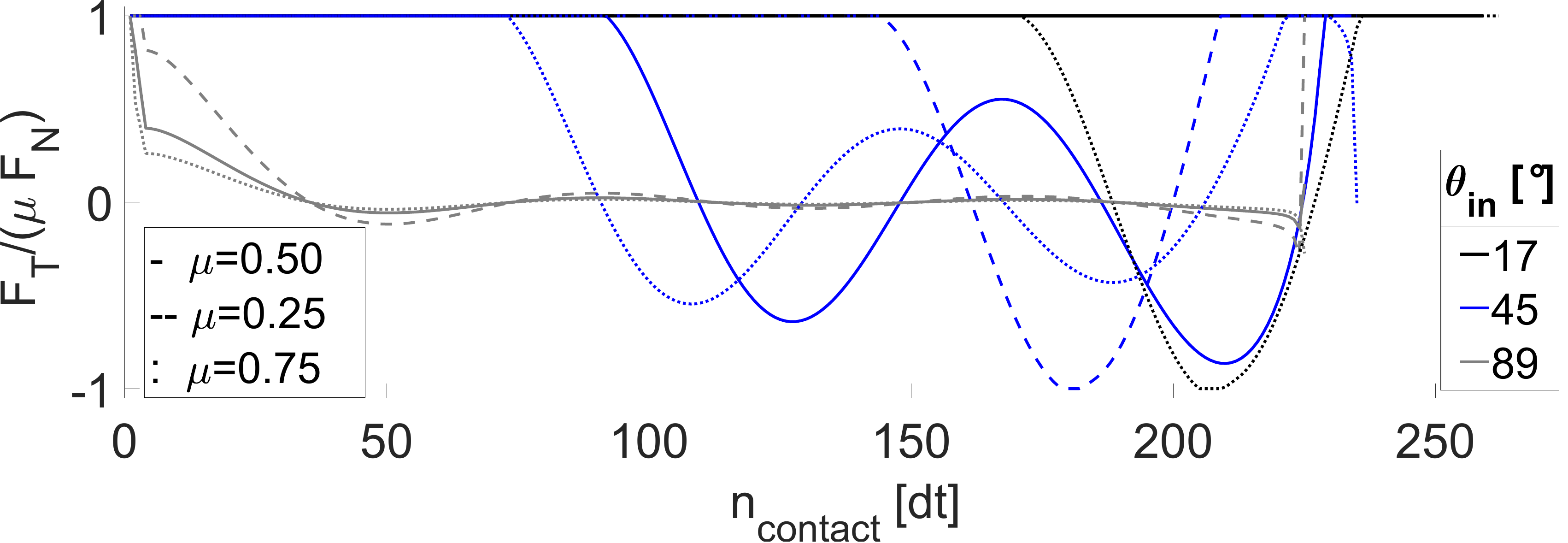}
\caption{Variation of the friction mobilization during the impact at different impact angles $\theta_{\mathrm{in}}$ and friction coefficients $\mu$.}
\label{fig:N120_contact_forcemobilization}
\end{figure}
The normal force also determines the maximum possible frictional response as $F_{T}=-\mu F_{N}$. Figure\,\ref{fig:N120_contact_forcemobilization} shows the evolution of the mobilized friction during the impact,
\begin{equation}
  -\frac{F_{T}}{\mu F_{N}},
\end{equation}
normalized by the applied coefficient of friction $\mu$. Following Stronge\,\cite{Stronge2000}, we find that at small angles of impact, the contact is continuously sliding throughout the impact, i.e. $\frac{|F_{T}|}{\mu F_{N}}=1$. The only exception occurs for large values of $\mu$, where towards the end of the contact, the contact enters the sticking regime $\frac{|F_{T}|}{\mu F_{N}}\leq 1$, before returning back to the sliding regime.
For steep impacts, regardless of the value of $\mu$, we observe sliding only during the initialisation of the contact, but the force ratio quickly decreases to zero. Larger values of $\mu$ lead only to a quicker decrease and smaller oscillation amplitudes around zero. For intermediate impact angles, the the contact is initially sliding, but later enters the sticking regime, with the onset stick being dependent on the magnitude of $\mu$. As for steep impact, the force ratio oscillates around zero decreases in the case of large $\mu$, but increases for $\mu\leq0.5$. In contrast to flat impacts, maximum inverse force response is reached at small values of $\mu$.

\begin{figure}[t]
\includegraphics[width=\columnwidth]{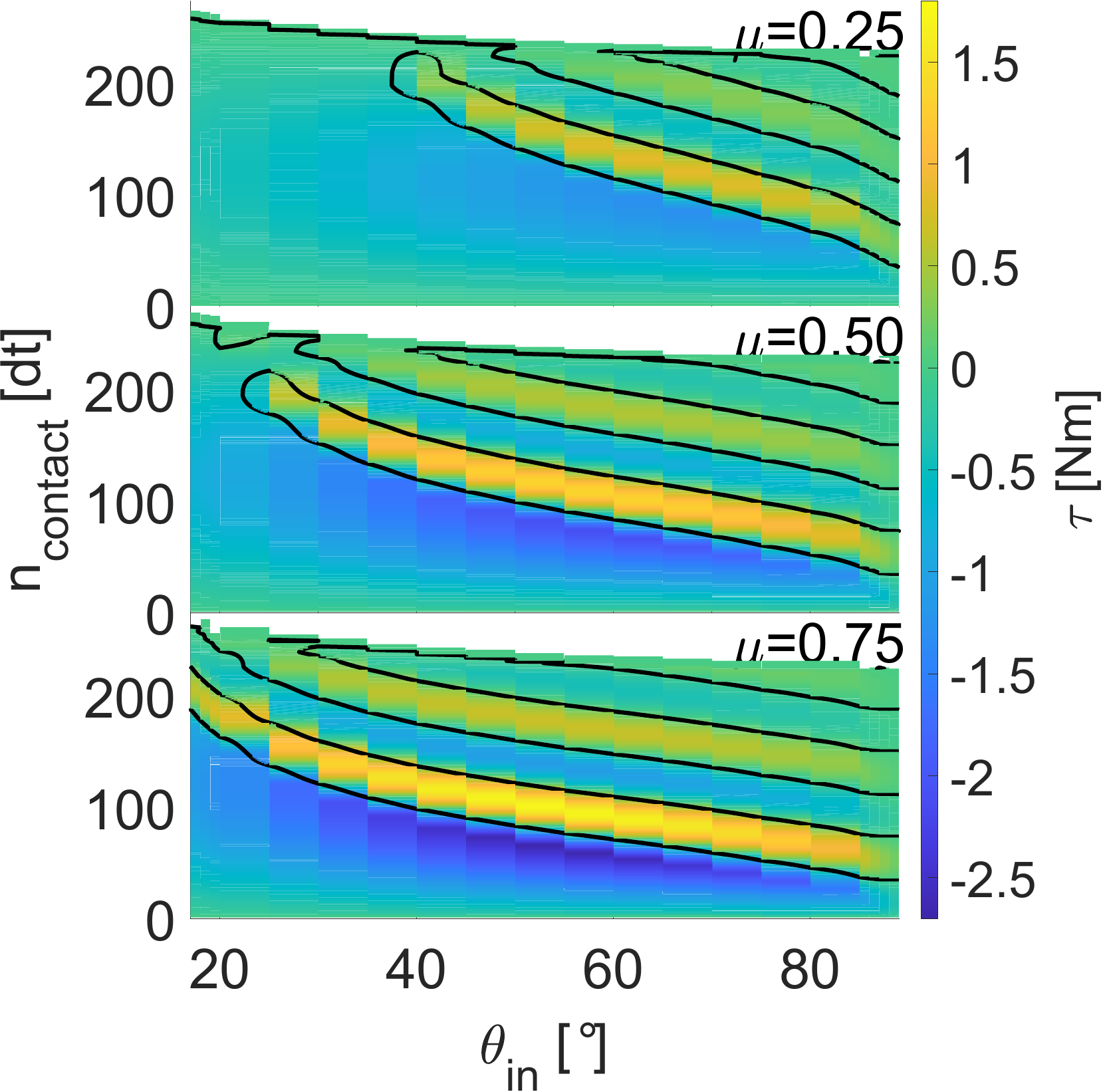}
\caption{Evolution of the torque $\tau$ acting on the disc during impact with the contact step $n_\mathrm{t}$ for different impact angles $\theta_{\mathrm{in}}$ and friction coefficients $\mu$.}
\label{fig:N120_contact_T}
\end{figure}

As the frictional forces induce torque, we further find the same wave pattern in the time evolution of the torque (Fig.\,\ref{fig:N120_contact_T}), where a strong negative torque first accelerates the particle into forward rotation ($\omega<0$), followed by the same deceleration / acceleration pattern with decaying magnitude shown previously for the tangential force. Since the induction of torque only depends on the presence of tangential forces, damping in normal direction does not affect the oscillations.

\begin{figure}[h]
\includegraphics[width=\columnwidth]{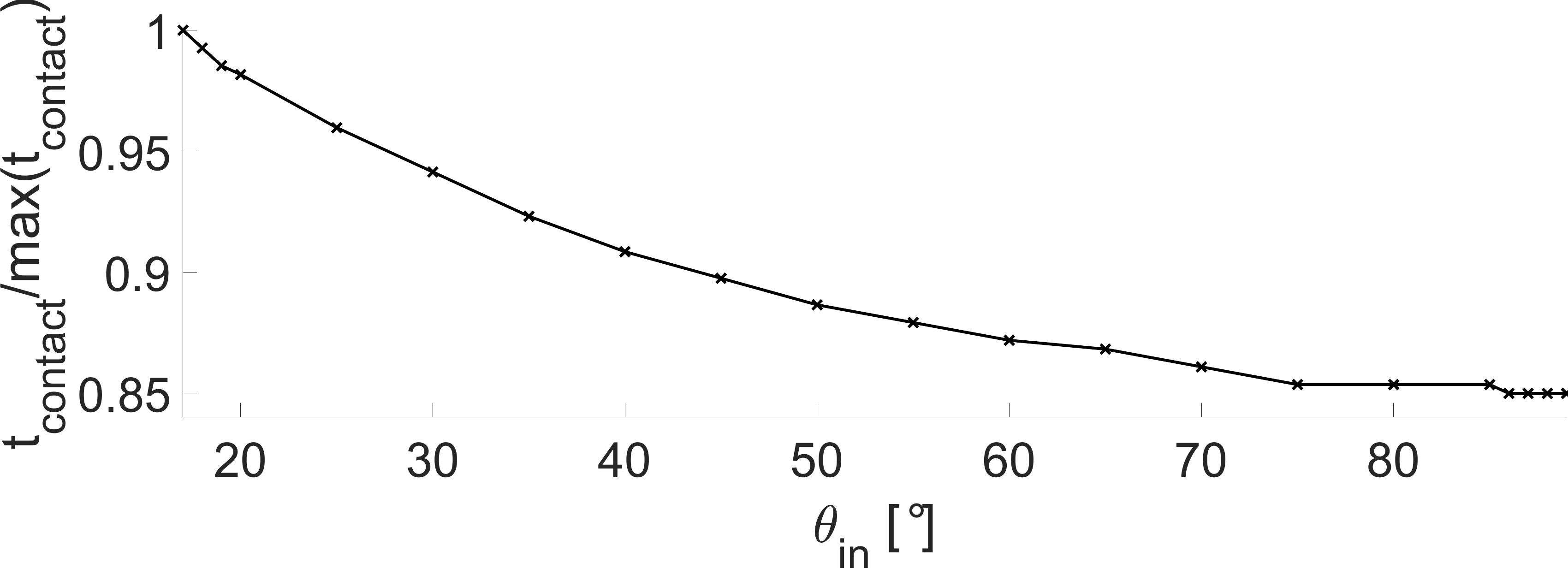}
\caption{Variation of the contact duration $t_{\mathrm{contact}}$ with impact angle $\theta_{\mathrm{in}}$.}
\label{fig:N120_contactduration}
\end{figure}

Lastly, we note that the duration of the contact decreases monotonically with increasing impact angle. At high impact angles, $t_{\mathrm{contact}}$ is up to 16\% shorter than for flat impacts, see Fig.\,\ref{fig:N120_contactduration}.

\section{Impact of a square particle}
As a disc is perfectly symmetric, there is no initial orientation to consider. On the other hand, the majority of granular particles are far from round. Their impact behaviour depends not only on their impact angle $\theta_{\mathrm{in}}$, but also their relative orientation upon contact\cite{Dattola2021}, and their motion during contact quickly becomes a complex combination of sliding and rolling\,\cite{Beunder2003,Zhao2018}. In this section we analyse the impact of a regular square onto a flat ground, with the same conditions as the disc in the previous section, but additionally with different initial orientations $\alpha$ of the particle. Here, $\alpha=0^{\circ}$ and $\alpha=90^{\circ}$ mean that a particle edge is initially parallel to the ground, while for $\alpha=45^{\circ}$ the centre of mass is vertically aligned with a vertex.

\subsection{Post-collision behaviour}
\begin{figure}[t]
\includegraphics[width=\columnwidth]{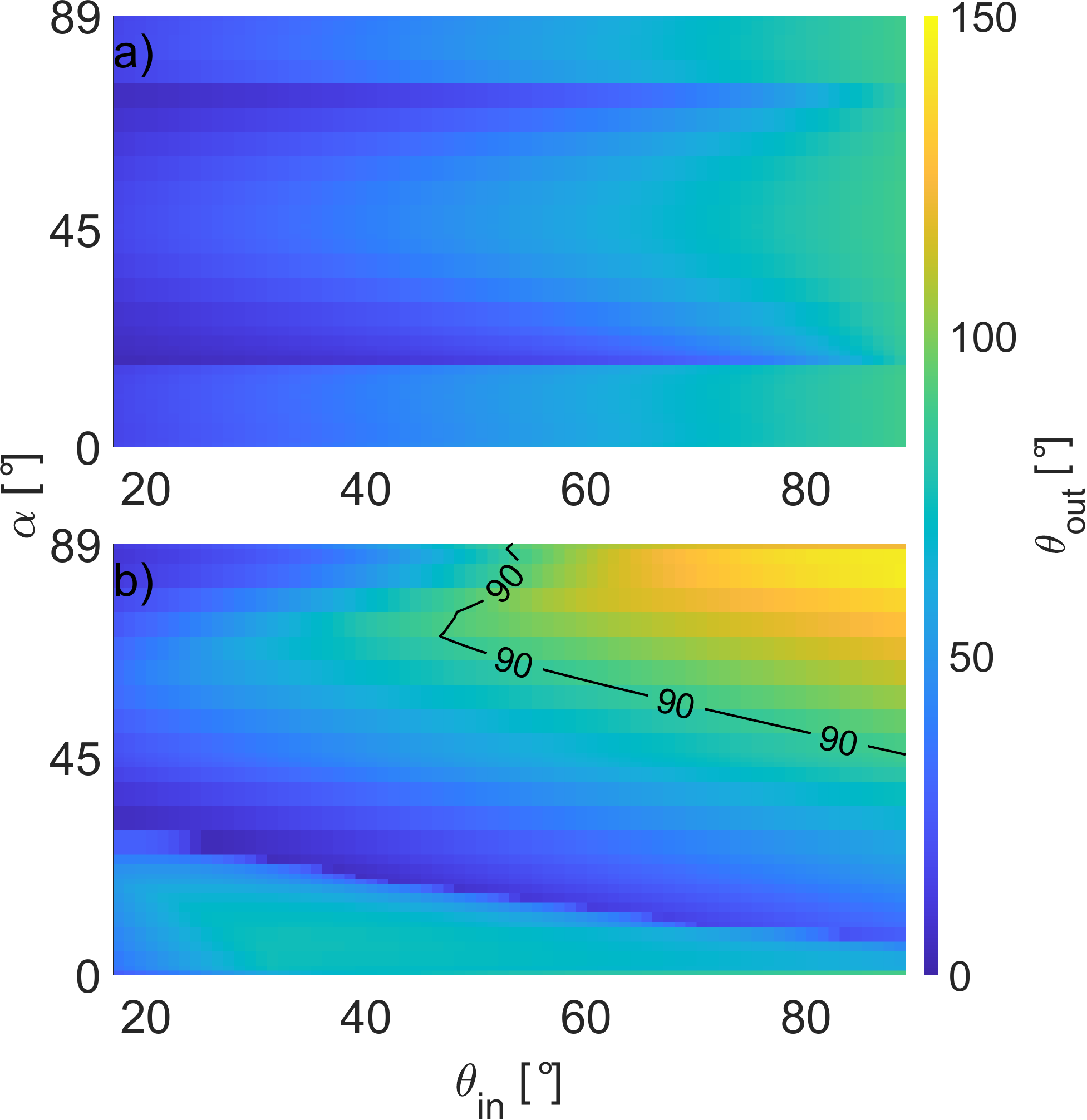}
\caption{Dependence of the rebound angle $\theta_{\mathrm{out}}$ on the combined influence of particle orientation $\alpha$ and impact angle $\theta_{\mathrm{in}}$ for the planar impact of square particles with a)$\tau \neq 0, \gamma=0.0,\mu=0.0$, and b) $\tau \neq 0, \gamma=0.5,\mu=0.5$. The encircled area in the top right corner of b) marks $\theta_{\mathrm{out}}>90^{\circ}$, i.e. backward deflection.}
\label{fig:N004_reboundangle}
\end{figure}

For square particles with inhibited rotation, the rebound angle $\theta_{\mathrm{out}}$ shows the same behaviour as for discs, see Fig\,\ref{fig:N004_reboundangle_norot} in the appendix. The rebound angle increases either equals the impact angle $\theta_{\mathrm{in}}$, or increases quickly before oscillating around $\theta_{\mathrm{out}}=90^{\circ}$. 
However, unlike for round particles, for square particles torque can also induced due to shape. Thus, if the particle is allowed to rotate, but experiences no dissipative forces ($\mu=0,\gamma=0$), the rebound angle changes with the initial particle orientation $\alpha$, see Fig.\,\ref{fig:N004_reboundangle}, a).  Going from $\alpha = 0^{\circ}\rightarrow\sim20^{\circ}$, $\theta_{\mathrm{out}}$ is independent of $\alpha$ and increases equivalent to $\theta_{\mathrm{in}}$. At $\alpha\sim20^{\circ}$ significantly decreases. $\theta_{\mathrm{out}}\ll\theta_{\mathrm{in}}$ until around $\theta_{\mathrm{in}}\sim80{\circ}$, where it rapidly increases to equivalence. For $\alpha \sim20^{\circ}\rightarrow\sim45^{\circ}$, $\theta_{\mathrm{out}}$ increases again, and its dependence on $\theta_{\mathrm{in}}$ approaches equivalence. The dependence on $\alpha$ is symmetric, centred around $\alpha=45^{\circ}$, as this determines the initial overlap and thus the normal force response. Thus $\alpha = 0^{\circ}\rightarrow 45^{\circ}$ is equal to $\alpha = 90^{\circ}\rightarrow 45^{\circ}$.

For dissipative particles with rotation, the rebound angle depends on the combination of $\theta_{\mathrm{in}}$ and $\alpha$. For backwards facing particles ($\alpha\gtrsim45^{\circ}$), strong backward deflection ($\theta_{\mathrm{out}}>90^{\circ}$) is possible at moderate to high impact angles ($\theta_{\mathrm{in}}\gtrsim45^{\circ}$), marked by the black line in Fig.\,\ref{fig:N004_reboundangle}, b). A second regime of high rebound angles ($\theta_{\mathrm{out}}\lesssim90^{\circ}$) exists for forward facing, near parallel particle alignment ($\alpha<30^{\circ}$) at all impact angles, visible in the bottom left corner of Fig.\,\ref{fig:N004_reboundangle}, b). The viable range of $\alpha$ with high deflection decreases as $\theta_{\mathrm{in}}$ increases. In-between these two areas the rebound angle is low, i.e. the particle is primarily moving forward instead of upward. We note that the rebound angle is particular low on the limit of the high-rebound angle zone for forward facing particle orientation.

\begin{figure}[t]
\includegraphics[width=\columnwidth]{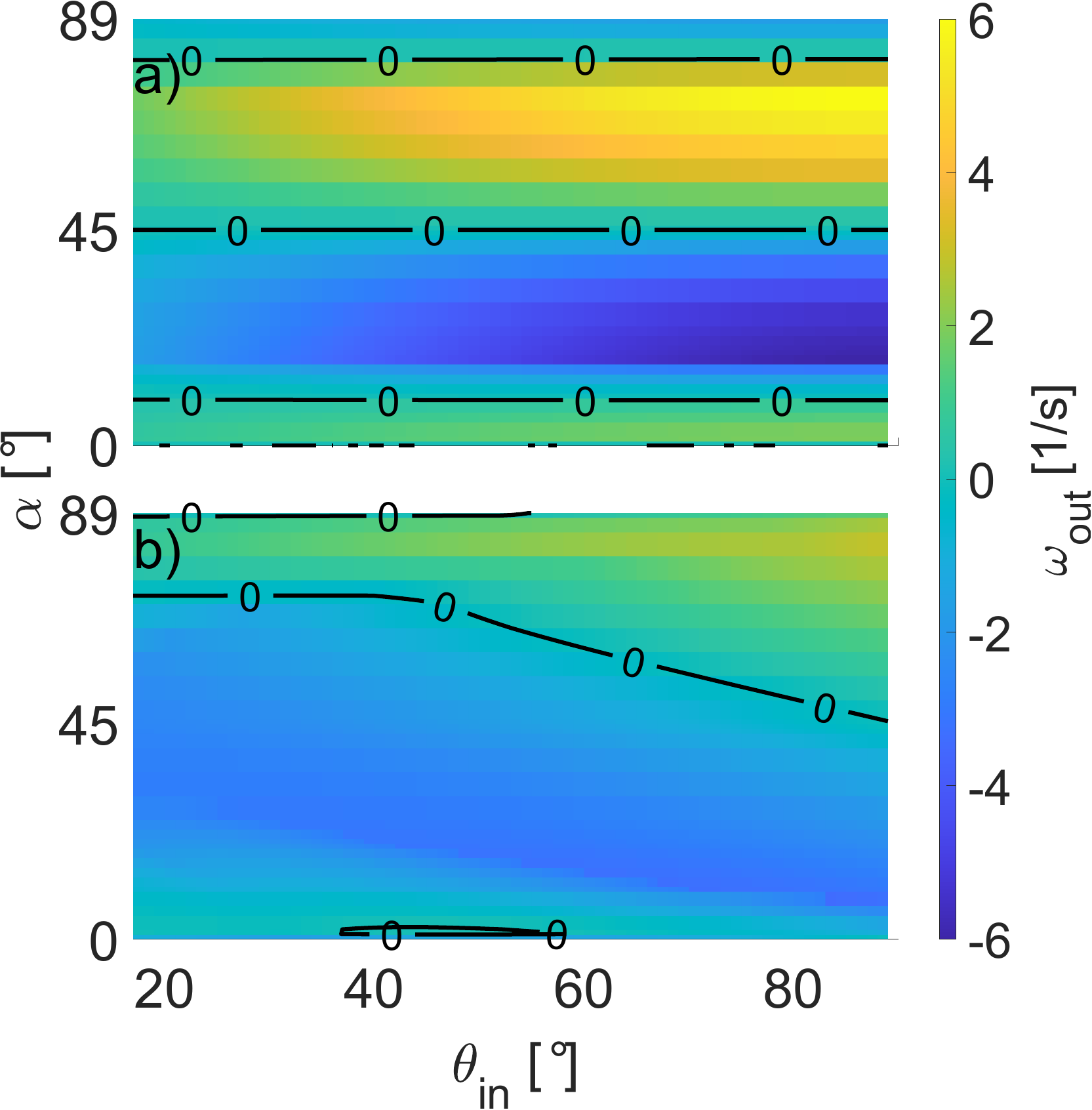}
\caption{Post-impact angular velocity for particles with rotational degrees of freedom as function of impact angle $\theta_{\mathrm{in}}$ and initial orientation $\alpha$. a) without dissipation ($\gamma=0.0,\mu=0.0$) and with dissipation ($\gamma=0.5,\mu=0.5$). The lines separate forward and backward rotation.}
\label{fig:N004_omega}
\end{figure}

The post-impact angular velocity $\omega_{\mathrm{out}}$ for non-dissipative particles (Fig.\,\ref{fig:N004_omega}, a) shows two zones of large rotation which match the zones of reduced rebound angle: A region of strong forward rotation ($\omega<0$) for forward oriented particles, and a region of strong backwards rotation ($\omega>0$) for backwards facing particles. The magnitude of rotation increases as the impact angle $\theta_{\mathrm{in}}$ grows larger. We associate both zones with the reduced rebound angle angle shown in Fig.\,\ref{fig:N004_reboundangle}, c). For near parallel particle orientation the angular velocity approaches zero, respectively is zero for parallel orientation.
For an impact with rotation and dissipative forces ($\gamma=0.5, \mu=0.5$, Fig.\,\ref{fig:N004_omega},b) we find a zone of strong backward rotation in the same ($\theta_{\mathrm{in}},\alpha$)-region in which backward deflection occurs. That the occurrence of backward rotation requires a steep impact angle and the particle facing backward (i.e. the centre of mass being behind the contact point) was also observed in experiment by Ji et al.\,\cite{Ji2023}.
In the second zone of high deflection angles in Fig.\,\ref{fig:N004_reboundangle},b) at $\alpha\lesssim20$, particles still rotate forward ($\omega<0$), however their angular velocity approaches zero. In the domain of low rebound angles, particles generally rotate forward with higher angular velocity ($\omega\rightarrow 0$). For backward facing, near parallel particles ($\alpha\gtrsim70$) and low impact angles ($\theta_{\mathrm{in}}\lesssim45^{\circ}$) we observe backward rotation, but the rebound angle is low ($\theta_{\mathrm{out}}\ll90$), i.e. the particle is moving forward.

\begin{figure}[t]
\includegraphics[width=\columnwidth]{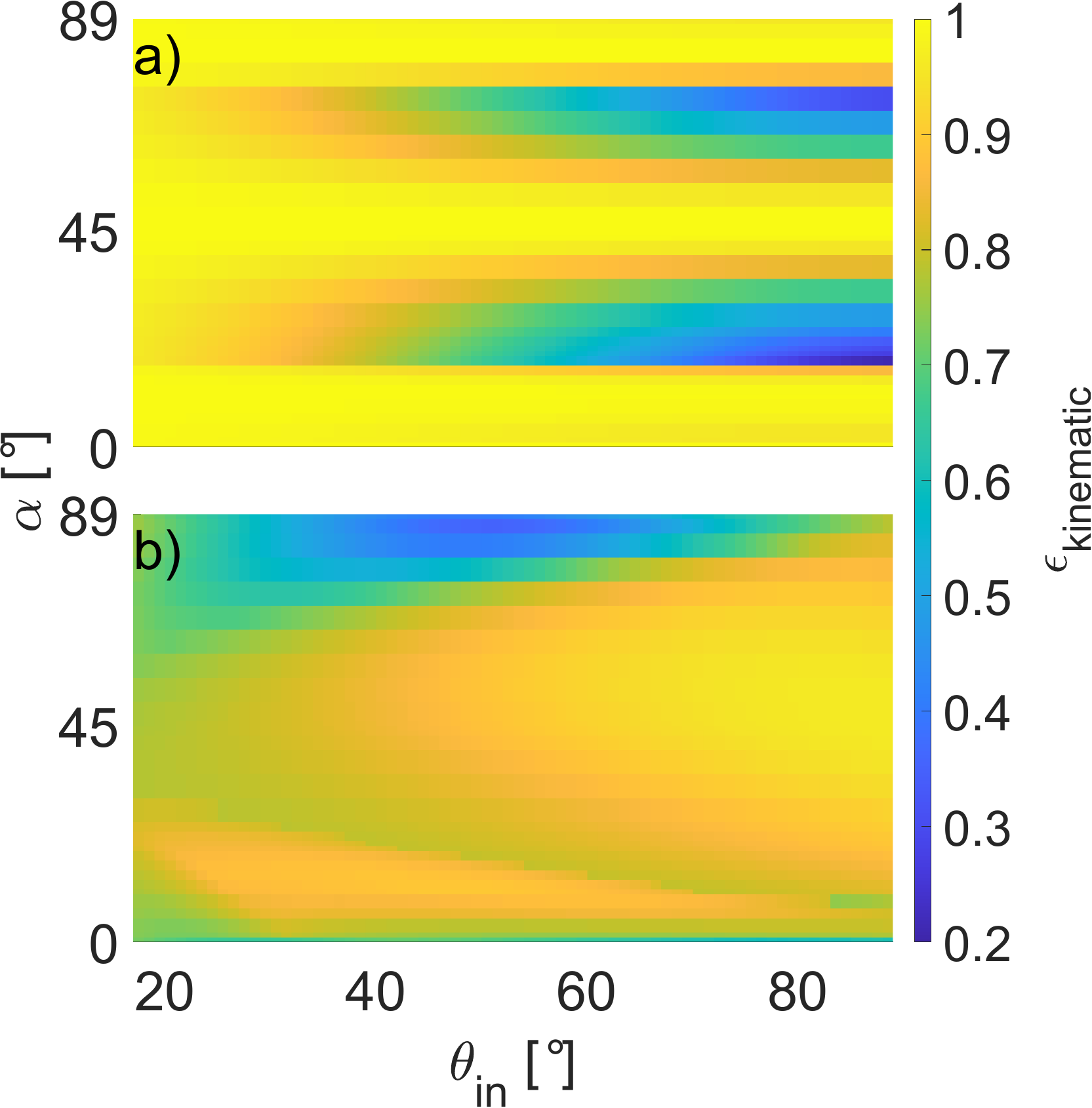}
\caption{Post-impact kinematic coefficient of restitution $\varepsilon_{\mathrm{kinematic}}$ for a) ($\gamma=0.0,\mu=0.0$), and b) ($\gamma=0.5,\mu=0.5$) with respect to impact angle $\theta_{\mathrm{in}}$ and initial orientation $\alpha$. For non-rotating, non-dissipative impacts, $\varepsilon_{\mathrm{kinematic}}=1=\mathrm{const}$.}
\label{fig:N004_COR_kinematic}
\end{figure}
In the case of non-dissipative particles without rotational degrees of freedom translational momentum is neither lost through dissipation, nor is it converted in the angular momentum, so the post-impact kinematic coefficient of restitution, eq.\,(\ref{eq:COR_kinematic}), remains $\varepsilon_{\mathrm{kinematic}} =  1 = \mathrm{const}$ for all impact angles and particle orientations.
For dissipative impacts without rotational degrees of freedom (Fig.\,\ref{fig:N004_cor_kinematic_norot} in the appendix), $\varepsilon_{\mathrm{kinematic}}$ behaves as it would for a disc (see Fig.\,\ref{fig:N120_restitution_kinematic} a): An initial decreases up to $\theta_{\mathrm{in}}\sim35^{\circ}$ is followed by an increase and saturation towards vertical impacts. As the particle cannot rotate in this case, the shape holds no influence, and the resulting kinematics are that of a disc with same circum-radius and material parameters. 
For non-dissipative impacts with rotation (Fig.\,\ref{fig:N004_COR_kinematic}, a), $\varepsilon_{\mathrm{kinematic}}$ mirrors the behaviour of the rebound angle $\theta_{\mathrm{out}}$ shown in Fig.\,\ref{fig:N004_reboundangle}, a). For near parallel orientation at impact, whether the particle is facing forward or backward, as well as for impacts with the centre of mass in vertical alignment with the lowest vertex, $\varepsilon_{\mathrm{kinematic}}\sim 1$ is largely conserved. In-between, for $\alpha \sim 30^{\circ}$ and $\alpha \sim 70^{\circ}$, $\varepsilon_{\mathrm{kinematic}}\sim 1$ for very flat impacts ($\theta_{\mathrm{in}}\lesssim30^{\circ}$), but decreases as $\theta_{\mathrm{in}}$ increases, corresponding to the increase in angular velocity shown in Fig.\,\ref{fig:N004_omega},a). 
Lastly, for impacts with both dissipation and rotation, $\varepsilon_{\mathrm{kinematic}}\gtrsim 0.8$ for almost all combinations of orientation $\alpha$ and impact angle $\theta_{\mathrm{in}}$. However, we also observe a significant decrease in $\varepsilon_{\mathrm{kinematic}}$ for $\alpha>70^{\circ}$ and $70^{\circ}<\theta_{\mathrm{in}}<90^{\circ}$. Symmetric with respect to particle orientation, we also find an area of slightly larger $\varepsilon_{\mathrm{in}}$, which matches the location of high rebound angles in Fig.\,\ref{fig:N004_reboundangle}, b) and the location of near zero angular velocity in Fig.\,\ref{fig:N004_omega},b).

\begin{figure}[t]
\includegraphics[width=\columnwidth]{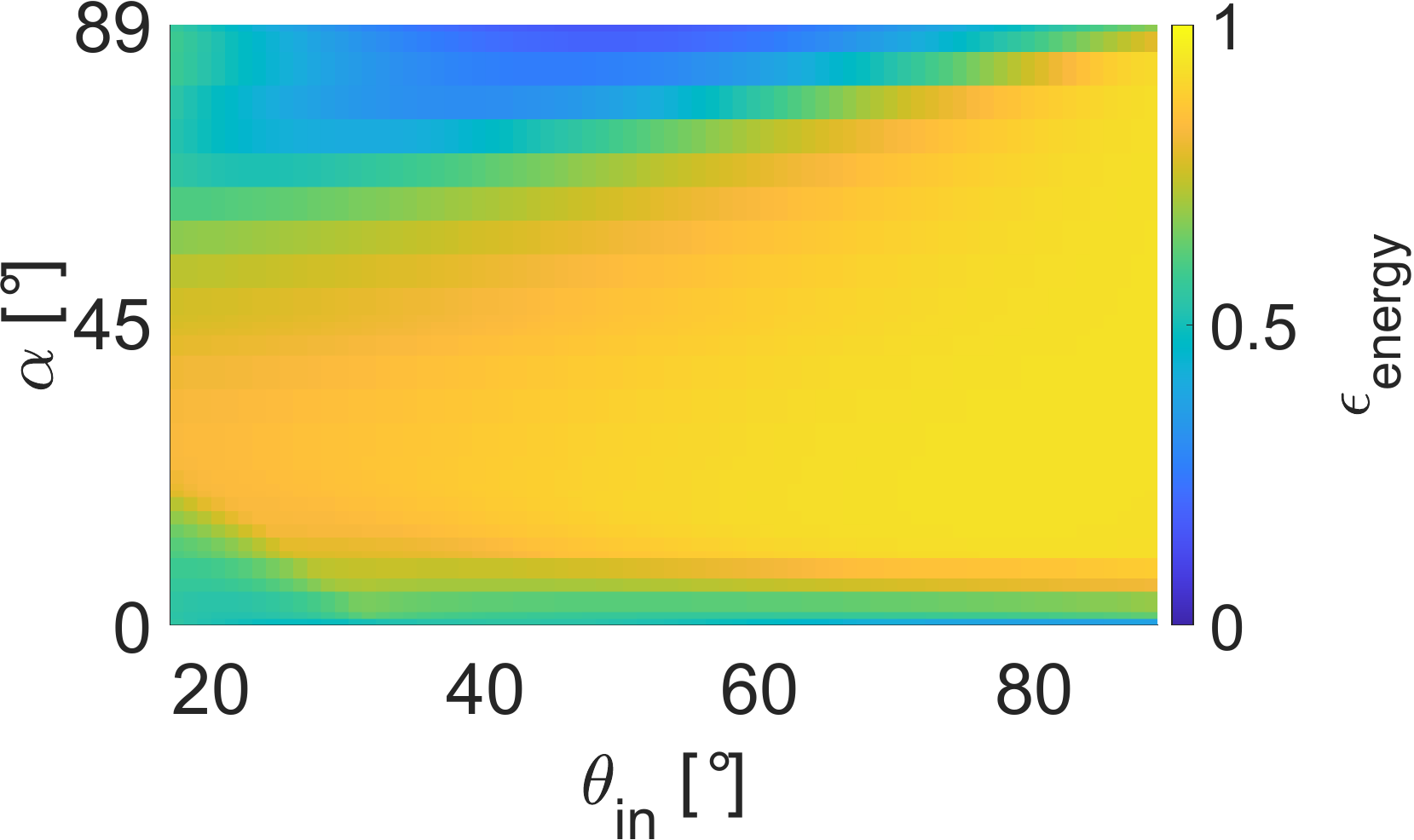}
\caption{Post-impact energy coefficient of restitution $\varepsilon_{\mathrm{energy}}$ for a dissipative particle with respect to impact angle $\theta_{\mathrm{in}}$ and initial orientation $\alpha$. Without dissipation $\varepsilon_{\mathrm{energy}}=1$ regardless of particle orientation or impact angle.}
\label{fig:N004_COR_energy}
\end{figure}

Besides the kinematic coefficient of restitution, eq.\,(\ref{eq:COR_kinematic}), other coefficients are commonly used, see e.g.\,\cite{Asteriou2015}.
The energy coefficient of restitution,
\begin{equation}
 \varepsilon_{\mathrm{energy}} = \frac{0.5mv_{\mathrm{out}}^2+0.5I\omega_{\mathrm{out}}^2}{0.5mv_{\mathrm{in}}^2+0.5I\omega_{\mathrm{in}}^2},
 \label{eq:COR_energy}
\end{equation}
gives a measure of the energy change of the particle including conversion of energy between translational to rotational degrees of freedom. As $\varepsilon_{\mathrm{energy}}$ considers both the rectilinear and rotational velocity of the particle it remains $\varepsilon_{\mathrm{energy}} = 1 = \mathrm{const.}$ in the absence of any dissipative force. For dissipative impacts without rotation (Fig.\,\ref{fig:N004_COR_energy_norot} in the appendix) $\varepsilon_{\mathrm{energy}}$ follows the functional form of the kinematic coefficient of restitution $\varepsilon_{\mathrm{kinematic}}$ although it decreases further, as the velocity magnitude is factored into the coefficient quadratically instead of linear. Similar, $\varepsilon_{\mathrm{energy}}$ for particles with both rotation and dissipation (Fig.\,\ref{fig:N004_COR_energy}) follows the kinematic coefficient of restitution, although the differences for $\varepsilon_{\mathrm{energy}}\lesssim 1$ vanish.
\begin{figure}[t]
\includegraphics[width=\columnwidth]{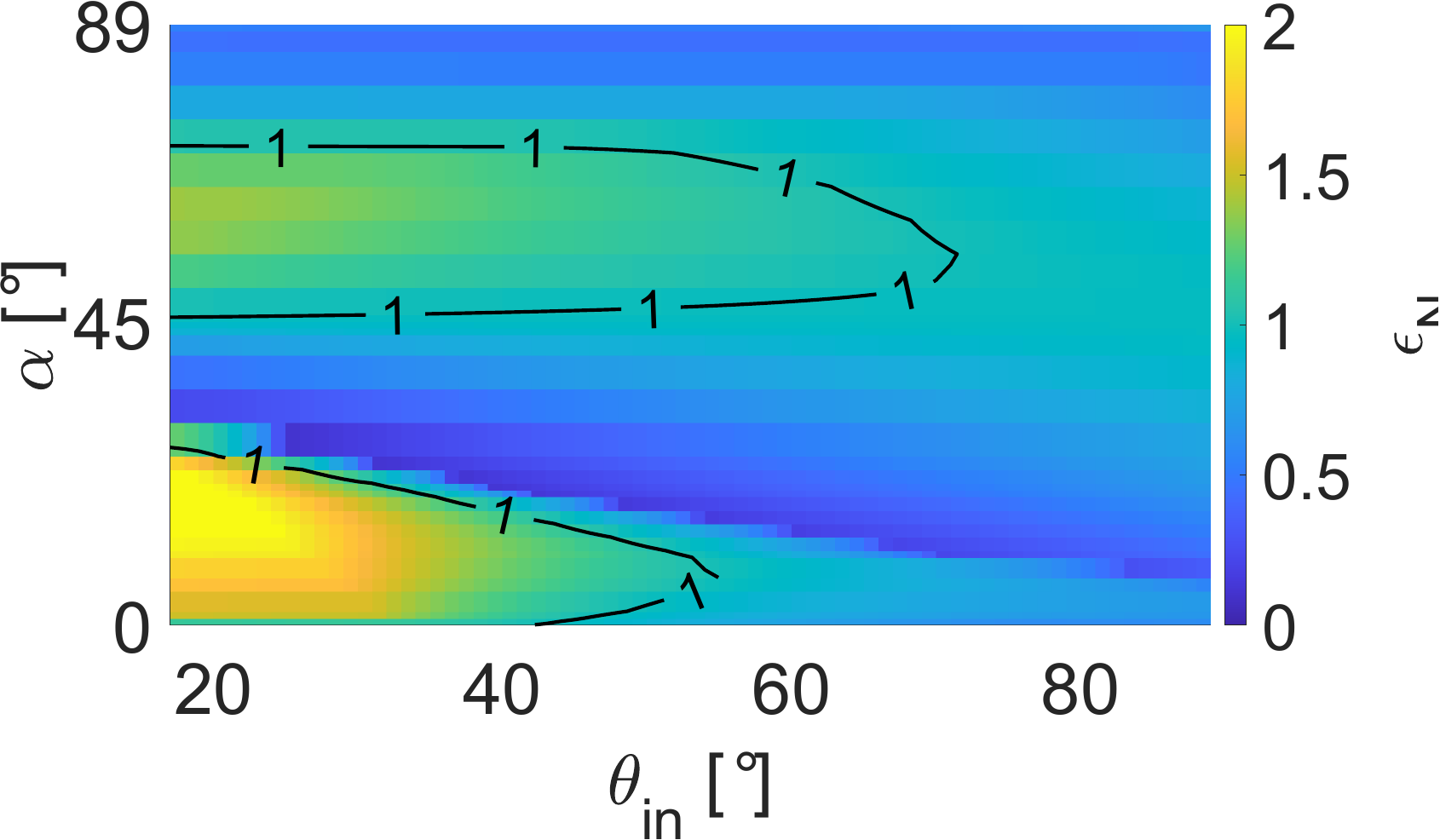}
\caption{Post-impact normal coefficient of restitution $\varepsilon_{\mathrm{N}}$ for ($\gamma=0.5,\mu=0.5$) with respect to impact angle $\theta_{\mathrm{in}}$ and initial orientation $\alpha$. Without rotational degrees of freedom and dissipative forces, $\varepsilon_{N}=1=\mathrm{const}$.}
\label{fig:N004_COR_normal}
\end{figure}

The normal coefficient of restitution,
\begin{equation}
 \varepsilon_{N} = -\frac{v_{y,\mathrm{out}}}{v_{y,\mathrm{in}}},
 \label{eq:COR_normal}
\end{equation}
is associated with the deformation of the particle upon contact. We find that for all $\alpha$ and $\theta_{\mathrm{in}}$, $\varepsilon_{N}=1=\mathrm{const}$, if the impact is either non-dissipative or if rotation is prohibited. For impacts with both dissipative forces and rotation we observe two zones with $\varepsilon_{N}>1$, for $\alpha\lesssim30^{\circ}$ and $\theta_{\mathrm{in}}\lesssim50^{\circ}$, and a second zone for $45^{\circ}\lesssim\alpha\lesssim70^{\circ}$ and $\theta_{\mathrm{in}}\lesssim70^{\circ}$. The first zone in particular shows a significant increase in $\varepsilon_{N}$ at low $\theta_{\mathrm{in}}$, indicating a large transfer of momentum from the tangential to the normal motion. This zone correlates with the zone of high rebound angles shown previously in Fig.\,\ref{fig:N004_reboundangle}, b).
The second zone at $\alpha\gtrsim 45^{\circ}$ however, does not correlates with either large rebound angles or backwards deflection. Outside of these two zones $\varepsilon_{N}$ quickly decreases to near zero values, partially corresponding to flat rebound angles $\theta_{\mathrm{out}}$, but also with backwards deflection. In contrast, we found $\varepsilon_{N}<1$ for impacts of discs at all impact angles. $\varepsilon_{N}>1$ also has been reported in previous numerical\,\cite{Saitoh2010,Mueller2012} and experimental studies\,cite{Wang2018,Fernandez2021} and is usually assumed to be $\varepsilon_{N}\in[0,1]$ in modelling of impact processes. However, $\varepsilon_{N}>1$ should not be taken as unusual, as it merely indicates energy transfer into the normal direction from any other degree of freedom (e.g. tangential motion, rotation, internal vibration), which is easily possible for non-spherical particles.

\begin{figure}[t]
\includegraphics[width=\columnwidth]{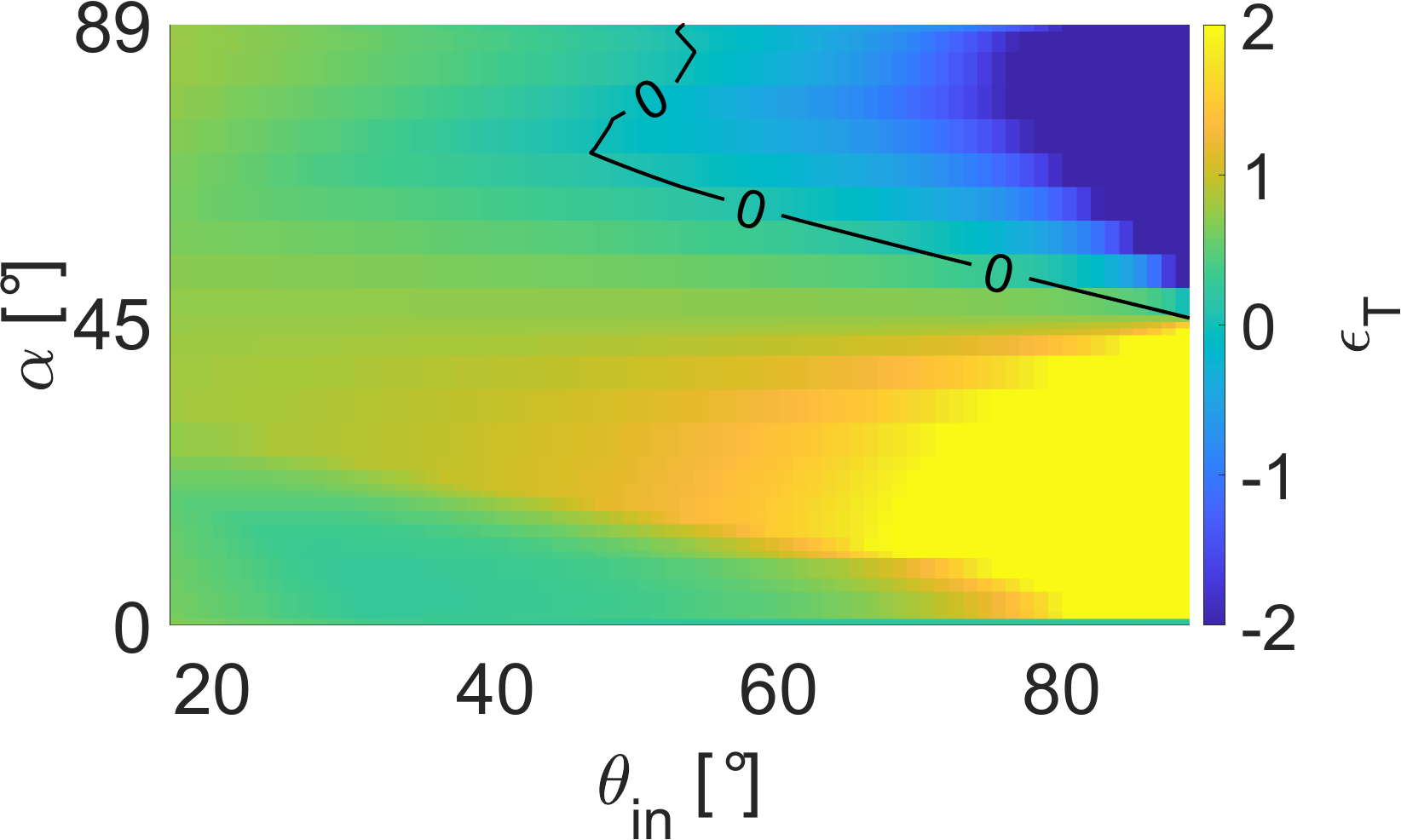}
\caption{Post-impact tangential coefficient of restitution $\varepsilon_{\mathrm{T}}$ for a) ($\tau=0,\gamma=0.5,\mu=0.5$) and b) ($\gamma=0.5,\mu=0.5$) with respect to impact angle $\theta_{\mathrm{in}}$ and initial orientation $\alpha$. Without dissipative forces, $\varepsilon_{T}=1=\mathrm{const}$.}
\label{fig:N004_COR_tangential}
\end{figure}

Equivalently, the tangential coefficients of restitution is defined as
\begin{equation}
 \varepsilon_{T} = -\frac{v_{x,\mathrm{out}}}{v_{x,\mathrm{in}}},
 \label{eq:COR_tangential}
\end{equation}
and describes the influence of friction and the transformation of translational to rotational moments, and vice versa. In the absence of tangential forces, the tangential coefficient of restitution, $\varepsilon_{T} = 1$, independent of the initial particle orientation. For dissipative particles without rotation, $\varepsilon_{T}$ also follows the same functional form as for discs, shown in Fig.\,\ref{fig:N120_restitution_normal}: An initial decrease, followed by oscillation around $\varepsilon_{T} = 0$, see Fig.\,\ref{fig:N004_COR_tangential_norot} in the appendix.
For dissipative systems with rotation $\varepsilon_{T}$ reflects the $\theta_{\mathrm{in}}$-$\alpha$ dependence of the rebound angle, see Fig.\,\ref{fig:N004_COR_tangential}. Large deflection angles are associated with $\varepsilon_{T}\gtrsim 0$, and tangential motion is converted into normal motion (Fig.\,\ref{fig:N004_COR_normal}). Backward deflection corresponds to strongly negative $\varepsilon_{T}\rightarrow-2$ and backward rotation ($\omega>0$ in Fig.\,\ref{fig:N004_omega}). For any other combination of $\alpha$ and $\theta_{\mathrm{in}}$ we observe $\varepsilon_{T}>0$, in particular for $\alpha\lesssim45^{\circ}$ and $\theta_{\mathrm{in}}\gtrsim70^{\circ}$, although the magnitude of positive $\varepsilon_{T}$ does not appear to correlate with the rebound behaviour or angular velocity.
\begin{figure}[b]
\includegraphics[width=\columnwidth]{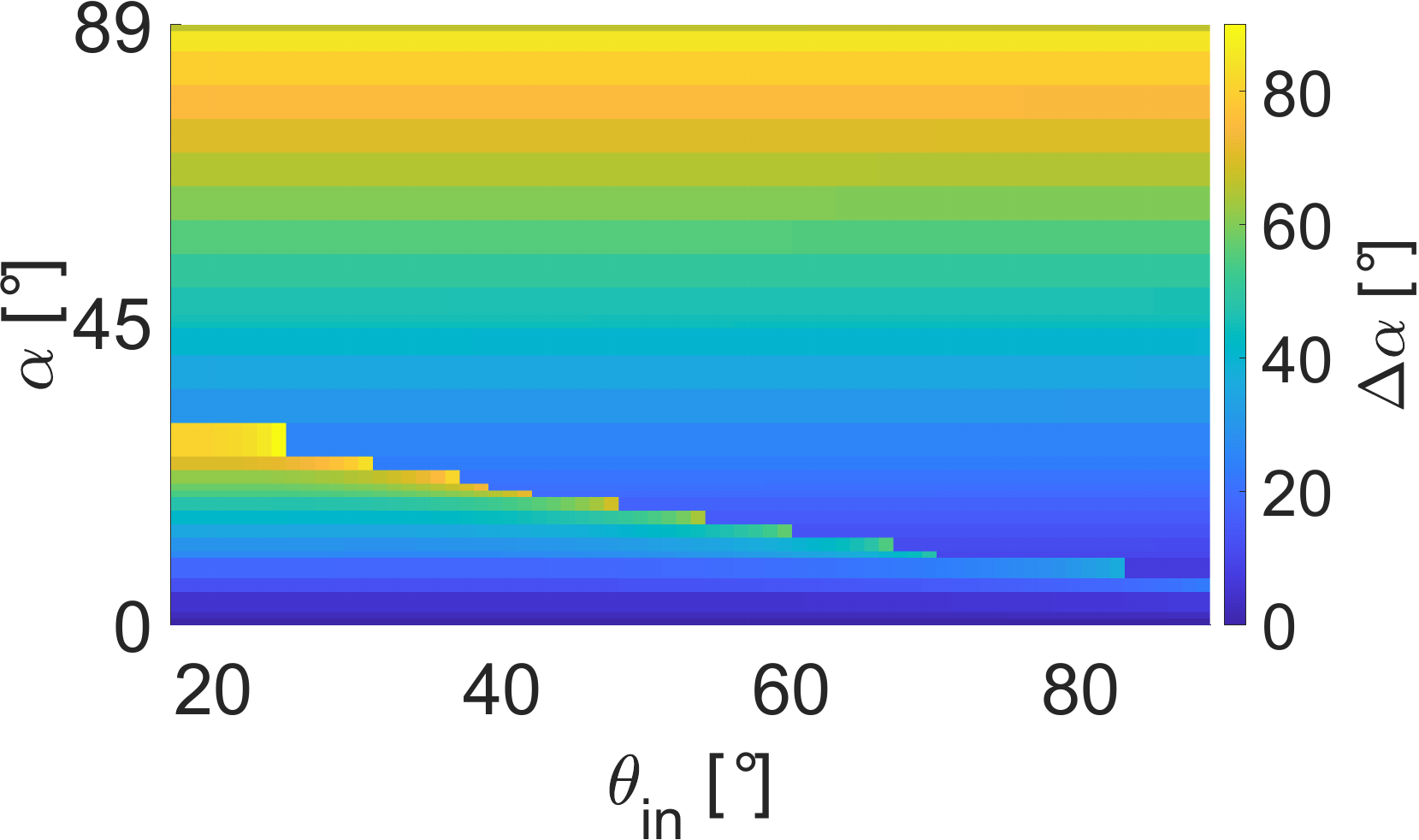}
\caption{Change in the particle orientation compared to initial orientation, $\Delta\alpha = \alpha_{\mathrm{out}}-\alpha_{\mathrm{in}}$, immediately upon detachment, for ($\gamma=0.5,\mu=0.5$) with respect to impact angle $\theta_{\mathrm{in}}$ and initial orientation $\alpha$.}
\label{fig:N004_alpha_out}
\end{figure}
As the particle orientation $\alpha$ upon contact plays an important role in shaping the collision behaviour of the particle, it is interesting to see how it changes after the collision. In Fig.\,\ref{fig:N004_alpha_out} we plot the difference between the initial particle orientation and the particle orientation upon detachment, $\Delta\alpha = \alpha_{\mathrm{out}}-\alpha$, with respect to both impact angle and initial particle orientation. For all combinations of $\alpha$ and $\theta_{\mathrm{in}}$ the particle is oriented backward after the contact. Further, for all impact angles, the orientation change generally becomes larger the more backward facing the particle is upon impact. However, we also note a line of strong impact changes crossing the plot diagonally, from $\alpha\sim30^{\circ}$ at $\theta_{\mathrm{in}}=17^{\circ}$ to $\alpha\lesssim10^{\circ}$ at $\theta_{\mathrm{in}}=89^{\circ}$. This zone roughly delineates the zone of high rebound angles and low angular velocity. Very little orientation change is obtained if the impact is either at near parallel, forward facing orientation upon impact, or terminates the contact with low angular velocity. The particle obtains angular momentum during the collision, so the orientation will keep changing during its free flight phase until the next impact.

\subsection{Behaviour during contact}
\begin{figure}[t!]
 \centering
 \includegraphics[width=0.95\columnwidth]{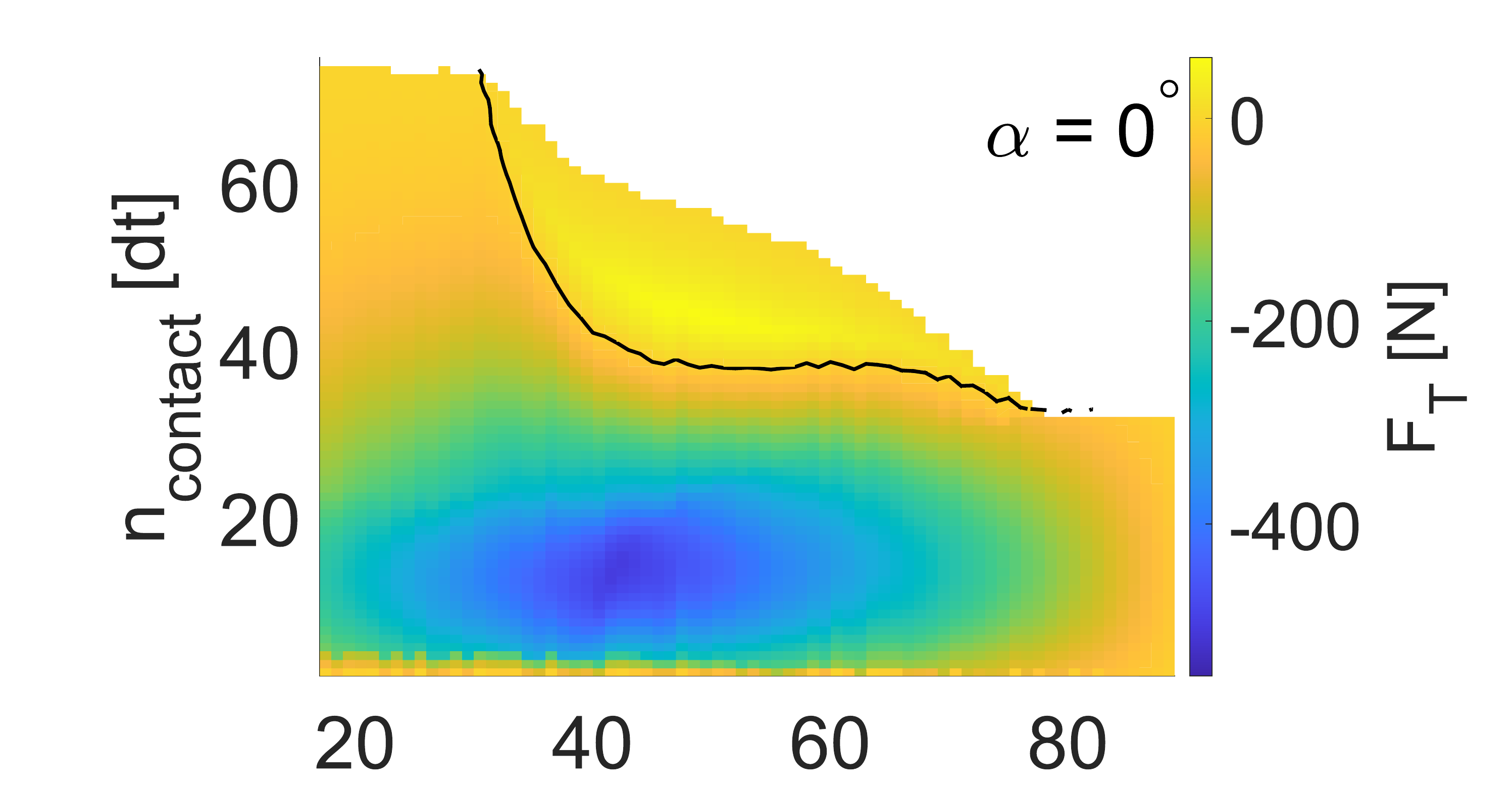}

 \includegraphics[width=0.95\columnwidth]{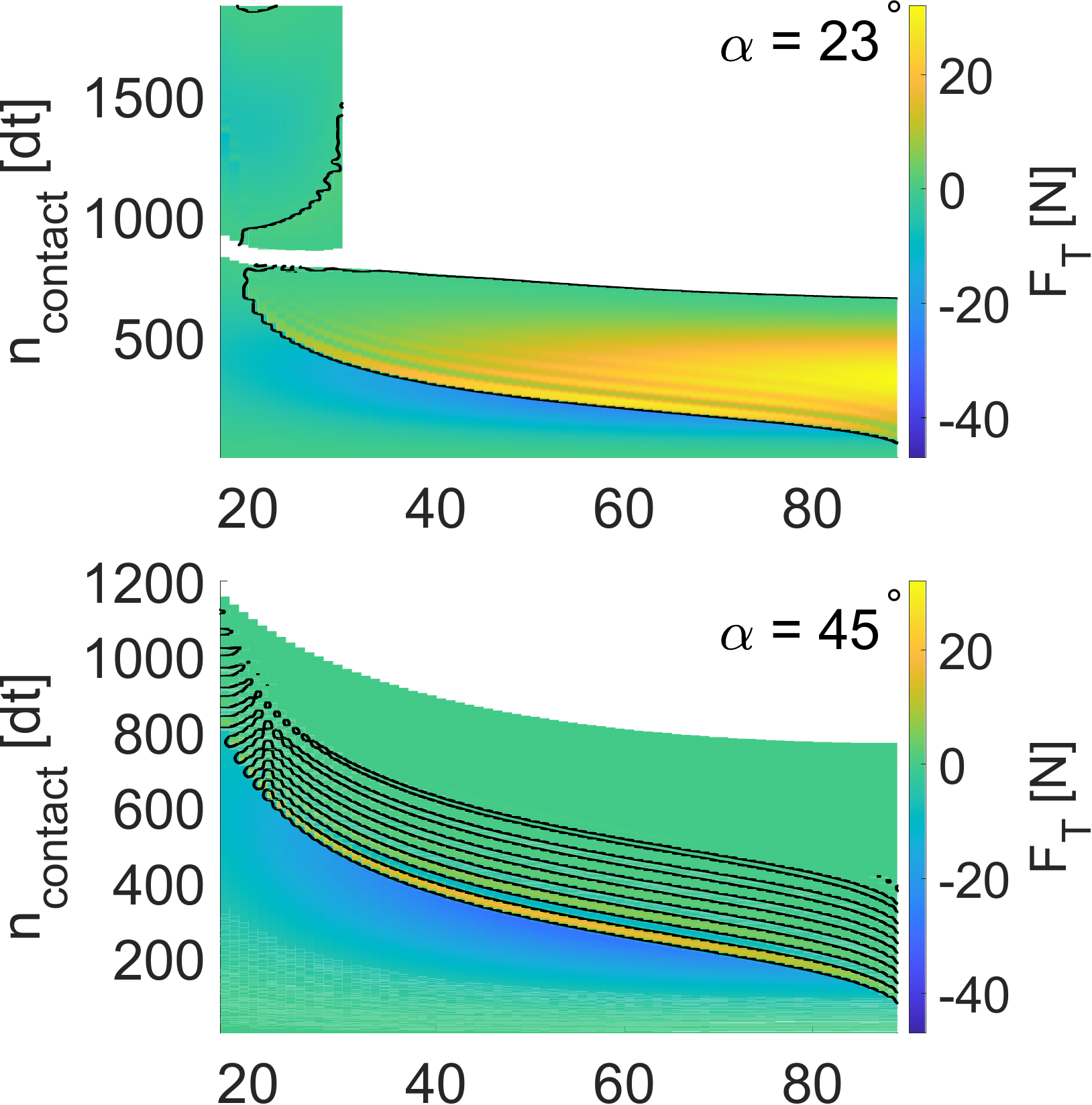}

 \includegraphics[width=0.95\columnwidth]{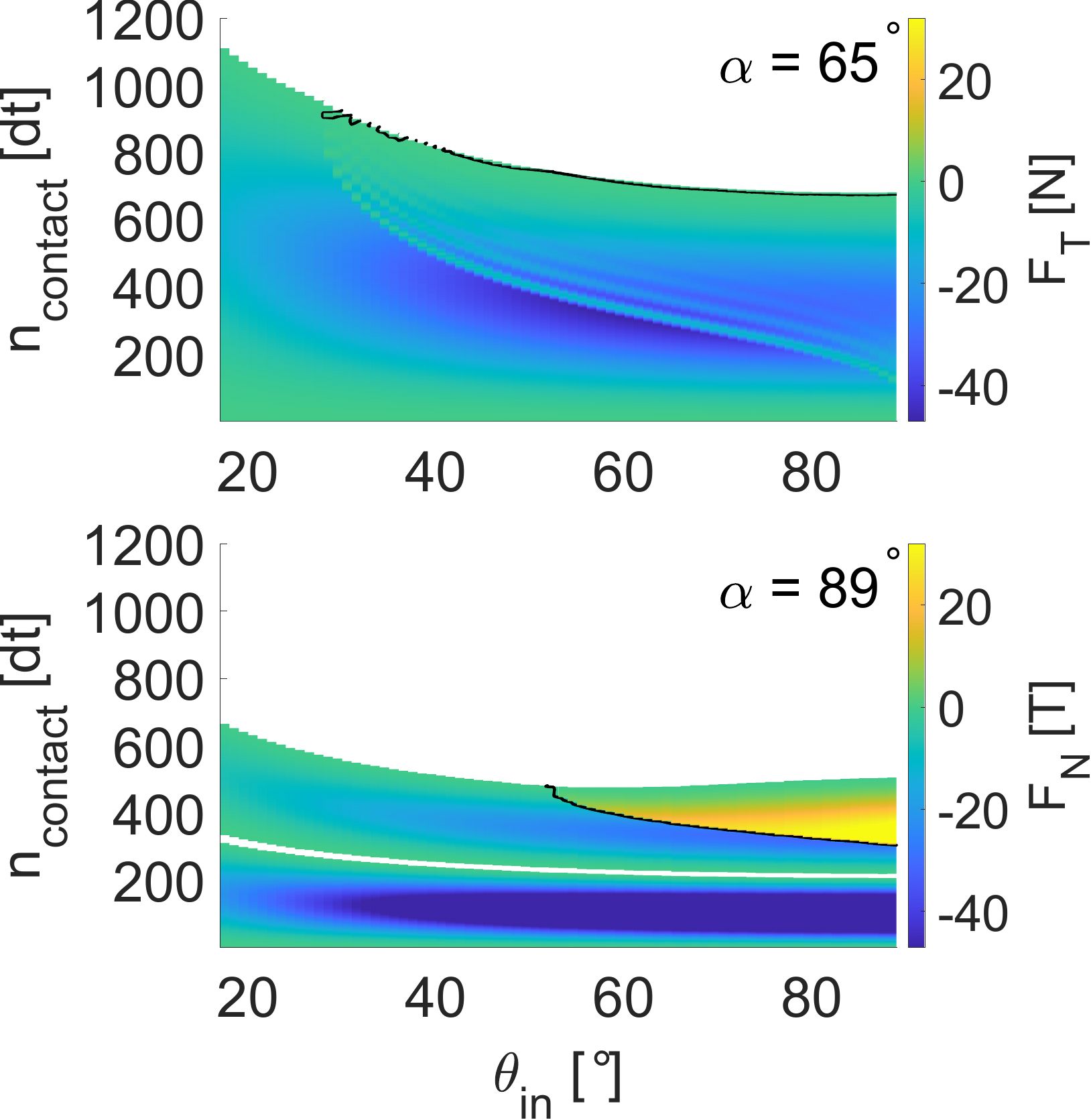}

 \caption{Evolution of the tangential force $F_{T}$ for ($\gamma=0.5,\mu=0.5$) during contact, for selected initial block orientations. Note that the colour-scale and contact duration change between different figures. The black line separates positive and negative acceleration.}
 \label{fig:N004_contact_fx}
\end{figure}

Unlike the disc, the initial particle orientation may significantly alter the contact evolution during the impact of a square. In the following we focus on evaluating the contact evolution for a square particle with rotational degrees of freedom and dissipative forces. 
We find that the evolution of the tangential contact force $F_{T}$ shows strong dependence on the particle orientation $\alpha$, see Fig.\,\ref{fig:N004_contact_fx}. For parallel orientation, $\alpha = 0^{\circ}$ the particle experiences primarily strong decelerating forces immediately after impact, in particular $\theta_{\mathrm{in}}\lesssim70^{\circ}$. Slight acceleration occurs at the end of the impact for $30^{\circ}\lesssim\theta_{\mathrm{in}}\lesssim75^{\circ}$. For non-parallel particle orientation ($\alpha>0^{\circ}$), the range of the experienced tangential forces decreases by an order of magnitude. 
As with the disc, we also observe wave-like patterns in $F_{T}$. For forward facing particles ($\alpha<45^{\circ}$) the force oscillates purely in the acceleration regime, but never decelerates after the oscillations begin.In contrast, when the particle faces backward during impact ($\alpha>45^{\circ}$) the force oscillates purely in the decelerating regime. If the particle centre is vertically aligned with its lowest vertex during contact ($\alpha=45^{\circ}$), we find a repeating pattern of acceleration and deceleration after the the initial deceleration, before the force converges towards zero and the contact terminates. 
We further note that the duration of the contact changes significantly with particle orientation: At $\alpha = 0^{\circ}$, the contacts are notably shorter than for any other initial particle orientation. 
For non-parallel particles, the contact duration increases by a factor of about 20, in particular for $\alpha<45^{\circ}$.

\begin{figure}[t]
\includegraphics[width=\columnwidth]{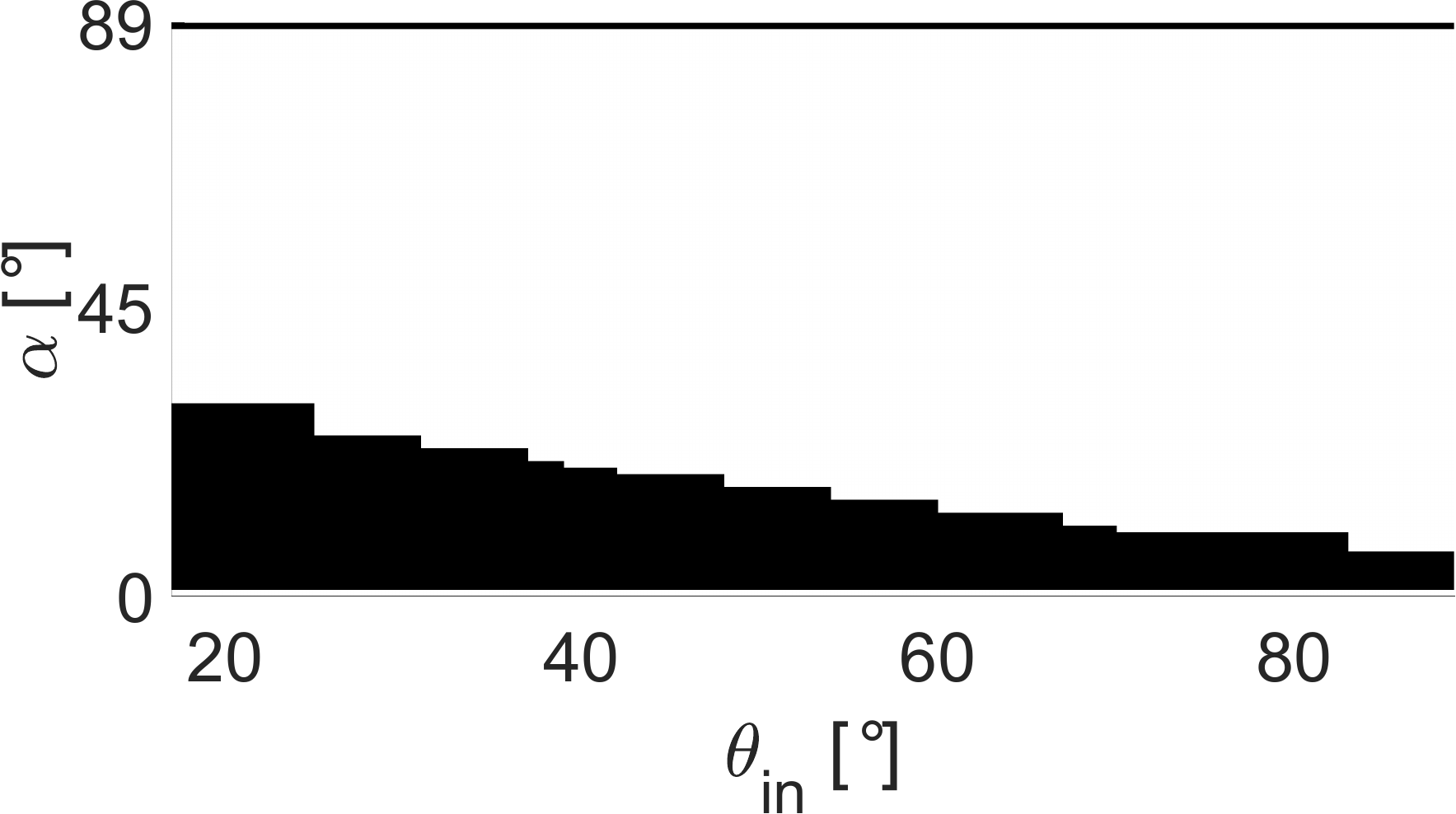}
\caption{Occurrence of double contacts (black) and single contacts (white) in the $\alpha$-$\theta_{\mathrm{in}}$ space.}
\label{fig:N004_doublecontact}
\end{figure}

Lastly, we find that for forward facing particles ($\alpha<45^{\circ}$) a ``double-contact'' may occur during a single impact event at certain impact angles, shown by the detached block for $\alpha=23^{\circ}$ in Fig.\,\ref{fig:N004_contact_fx}. In the second contact phase, the induced contact forces generally are much weaker than in the first contact phase before. A second ``double-contact'' occurs at $\alpha\sim89^{\circ}$. The ``double-contact'' at large $\alpha$ differs from the ``double-impact'' at small $\alpha$, in that the second contact phase shows stronger peak deceleration, but also allows stronger acceleration for $\theta_{\mathrm{in}}\gtrsim 50^{\circ}$.
The $\alpha$-$\theta_{\mathrm{in}}$-range in which this ``double-contact'' occurs, shown in Fig.\,\ref{fig:N004_doublecontact}, coincides with the zone of high deflection angles ($\theta_{\mathrm{out}}\rightarrow 90^{\circ}$) (Fig.\,\ref{fig:N004_reboundangle}, b), but not with backwards deflection $\theta_{\mathrm{out}}>90^{\circ}$. We can also associate it further with small angular velocity upon separation (Fig.\,\ref{fig:N004_omega}).

\begin{figure}[b]
\includegraphics[width=\columnwidth]{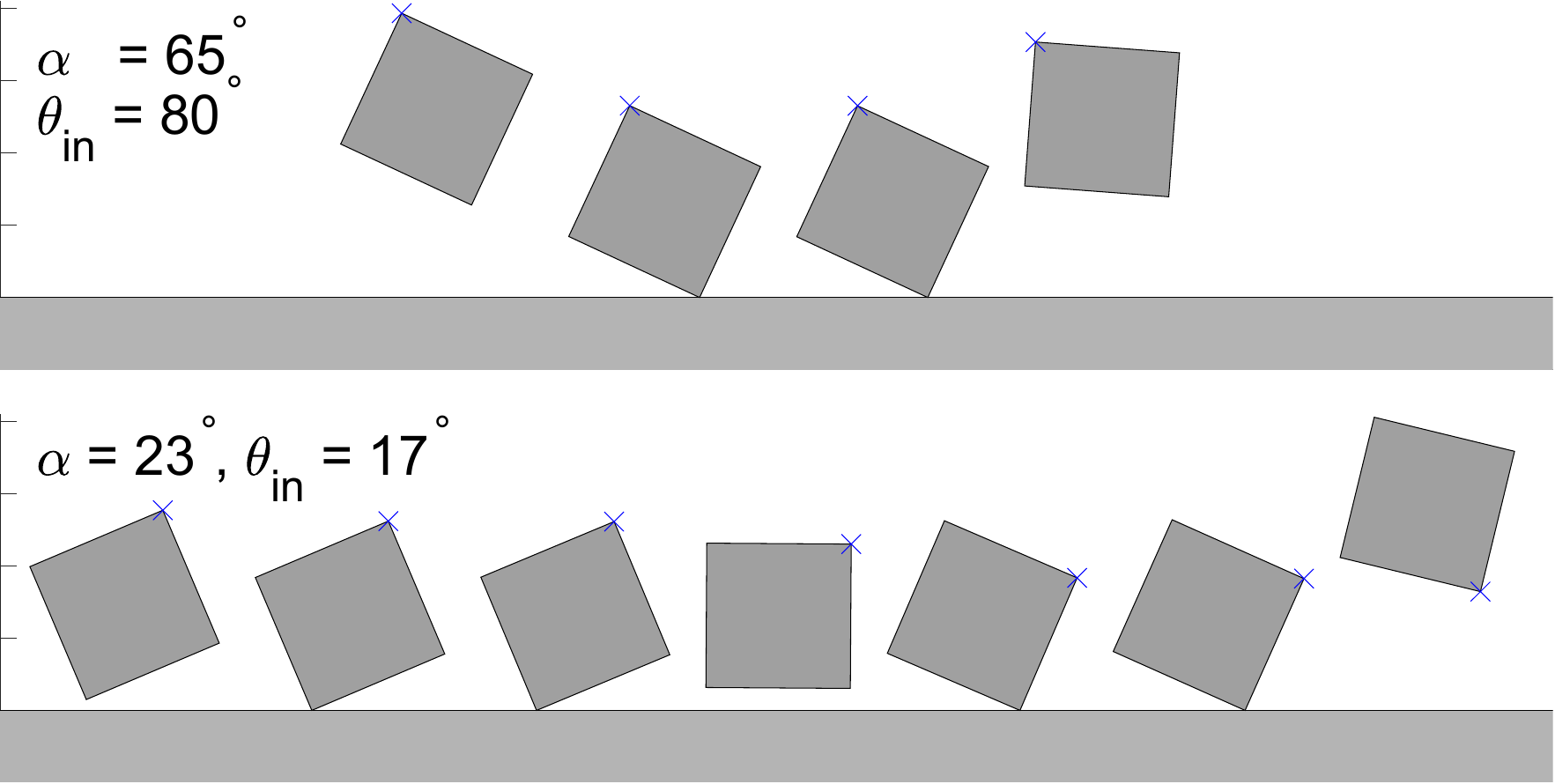}
\caption{Visualisation of the particle during impact for the steep impact of a backwards facing particle ($\alpha=64^{\circ},\theta_{\mathrm{in}}=80^{\circ}$) and for the ``double-contact'' of a forward facing particle during a flat impact ($\alpha=23^{\circ},\theta_{\mathrm{in}}=17^{\circ}$). The horizontal spacing of the particles is significantly exaggerated to improve readability of the figure. The blue x marks the location of the same corner of the block during the impact process.}
\label{fig:N004_contactvisual}
\end{figure}

Large changes in the particle orientation imply a large angular velocity of the particle, however the post-collision angular velocity for these particular impacts approaches zero. We therefore understand the origin of the high deflection angles as follows: During the first impact event, a large amount of translational velocity is converted into rotation. As the impact duration is short, the orientation of the particle changes only by a negligible amount, the same behaviour as for other particle orientations and impact angles. After detachment, the particle begins to rotate in forward direction ($\omega<0$). The normal rebound velocity has decreases significantly, so a vertex of the particle comes into contact with the plane before the particle has moved away. This initiates a second impact event with an initial angular velocity and a translational velocity vector that is pointing away from the plane. Now, energy is converted back from the rotational degree of freedom into the rectilinear degree of freedom, which leads to an increase in normal velocity over tangential velocity, and thus high deflection angles. In other cases, the particle either moves away from the plane before the lowest vertex can make contact or rotates in backward direction, so that a second impact event cannot be initiated. Figure.\,\ref{fig:N004_contactvisual} visualises the particle contact process for the two cases. Thus, during the second impact event, the particle is moving away from the plane upon initiating contact, instead of moving towards it.

\begin{figure}[t!]
 \centering
 \includegraphics[width=0.95\columnwidth]{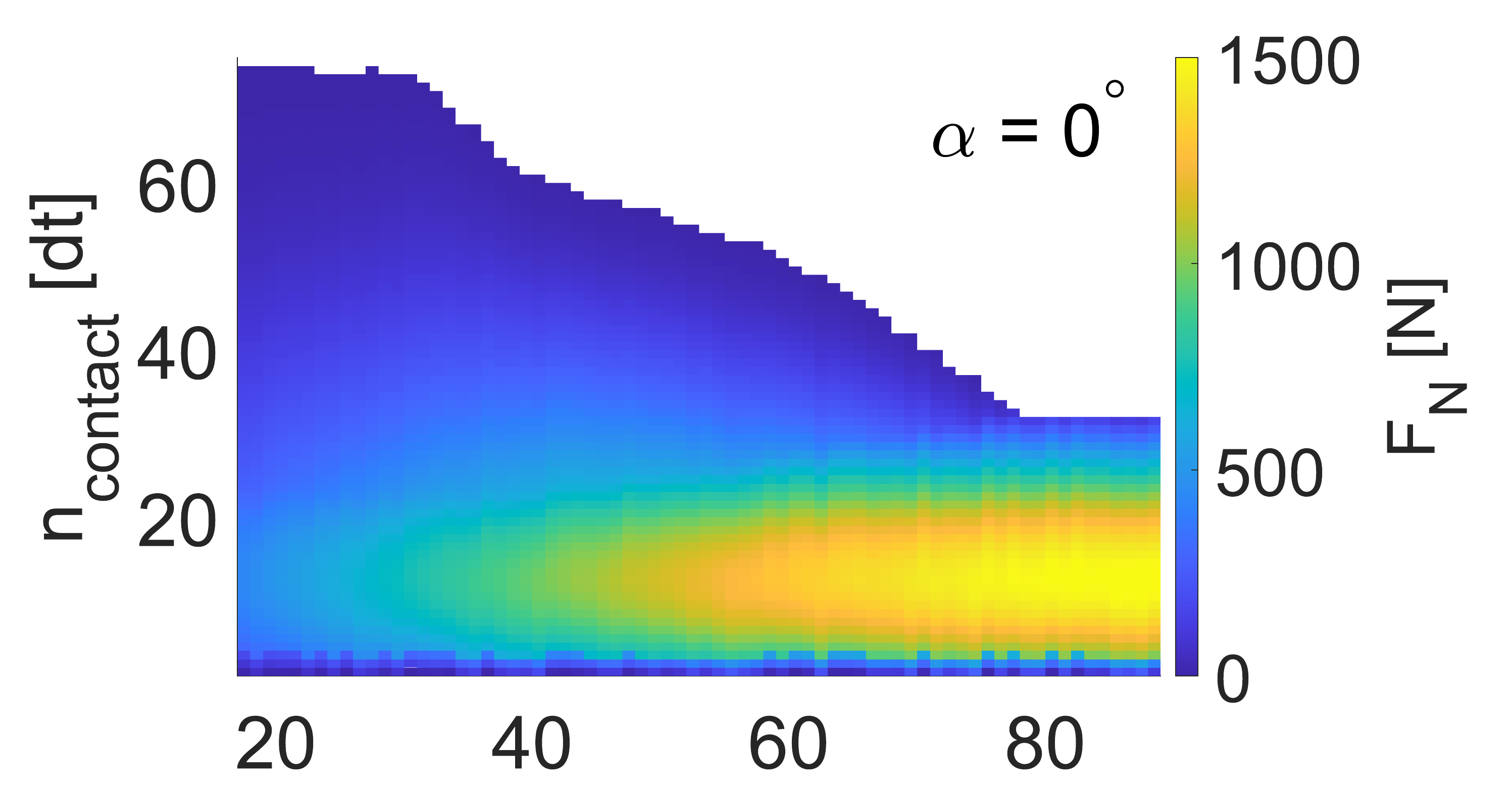}
 
 \includegraphics[width=0.95\columnwidth]{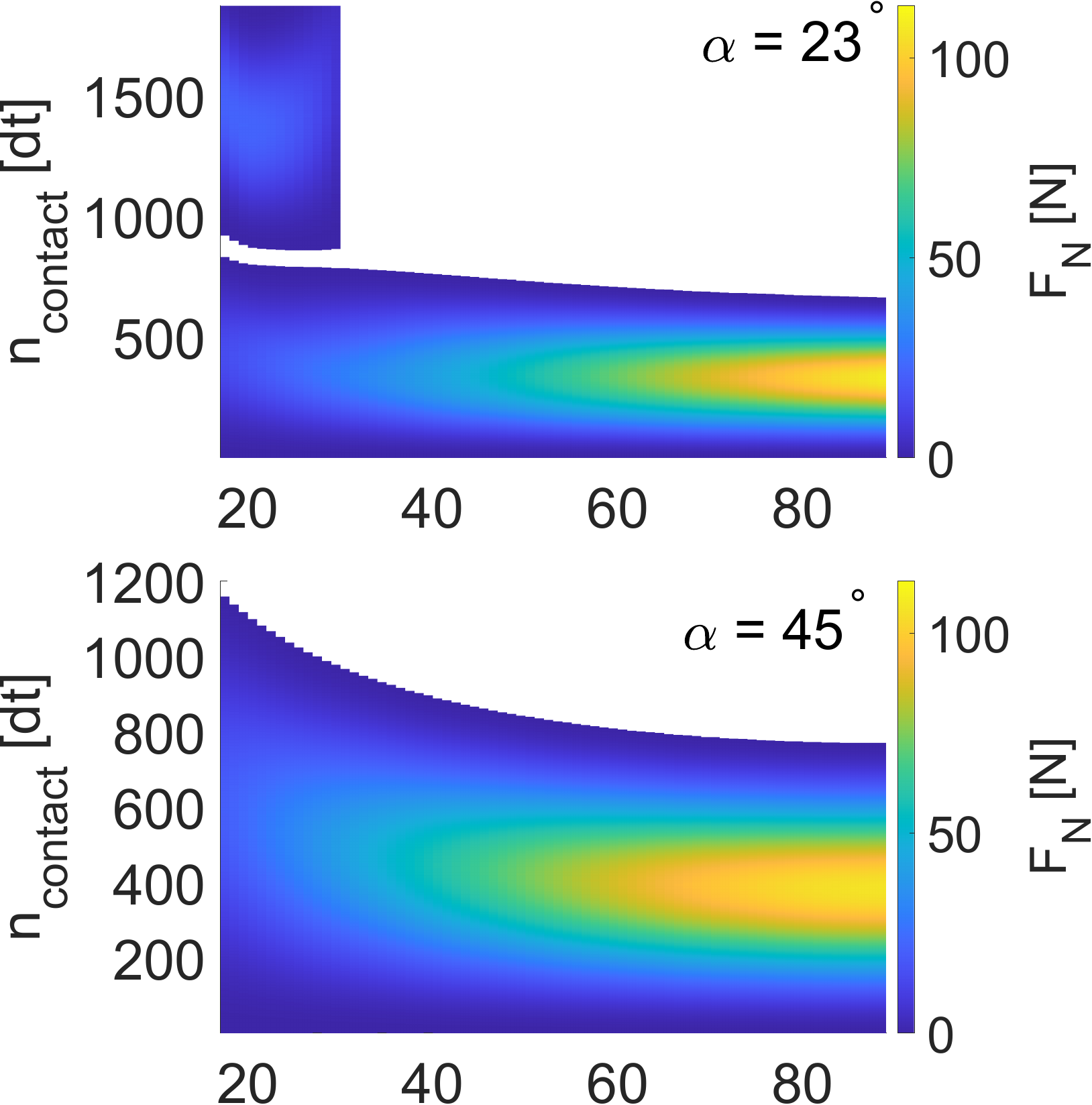}

 \includegraphics[width=0.95\columnwidth]{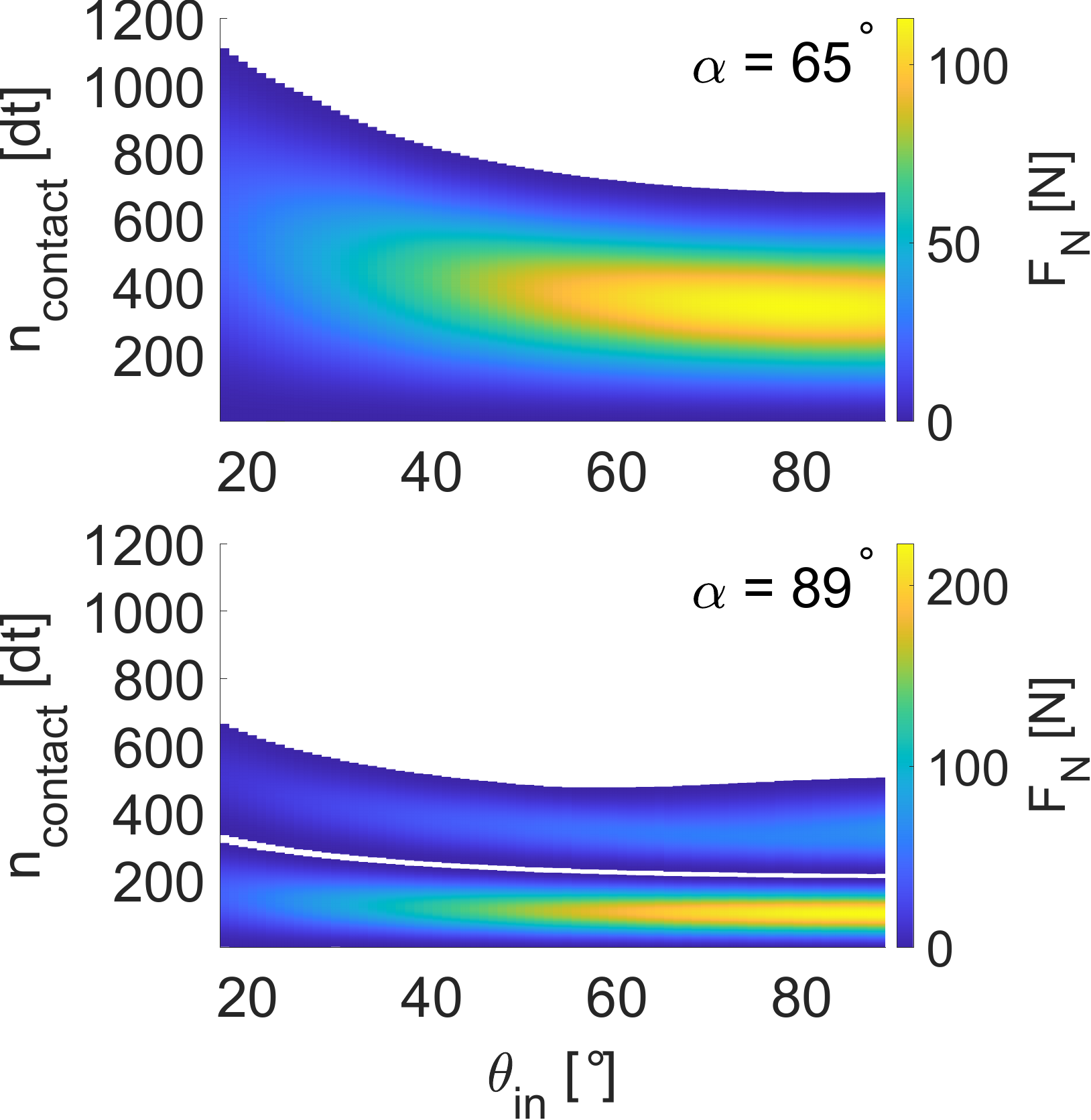}

 \caption{Evolution of the normal force $F_{N}$ for ($\gamma=0.5,\mu=0.5$) during contact, for selected initial block orientations. Note that the colour-scale and contact duration change between different figures.}
 \label{fig:N004_contact_fy}
\end{figure}
Like the tangential contact force, the magnitude of the contact normal force, $F_{N}$ for parallel-oriented particle impacts is significantly higher than for any other particle orientation: as the interface during contact is significantly larger than for tilted particles, more momentum can be transferred simultaneously. As a result, the induced forces are larger and the initial velocity is quicker reversed, which explains the much shorter duration of the contacts. As with the disc, the induced normal force is larger for steeper impacts, regardless of the initial particle orientation, as the normal component of the impact velocity is also much larger. In the second contact phase, for forward facing particles only, a much weaker normal force is induced.

\begin{figure}[t]
\includegraphics[width=\columnwidth]{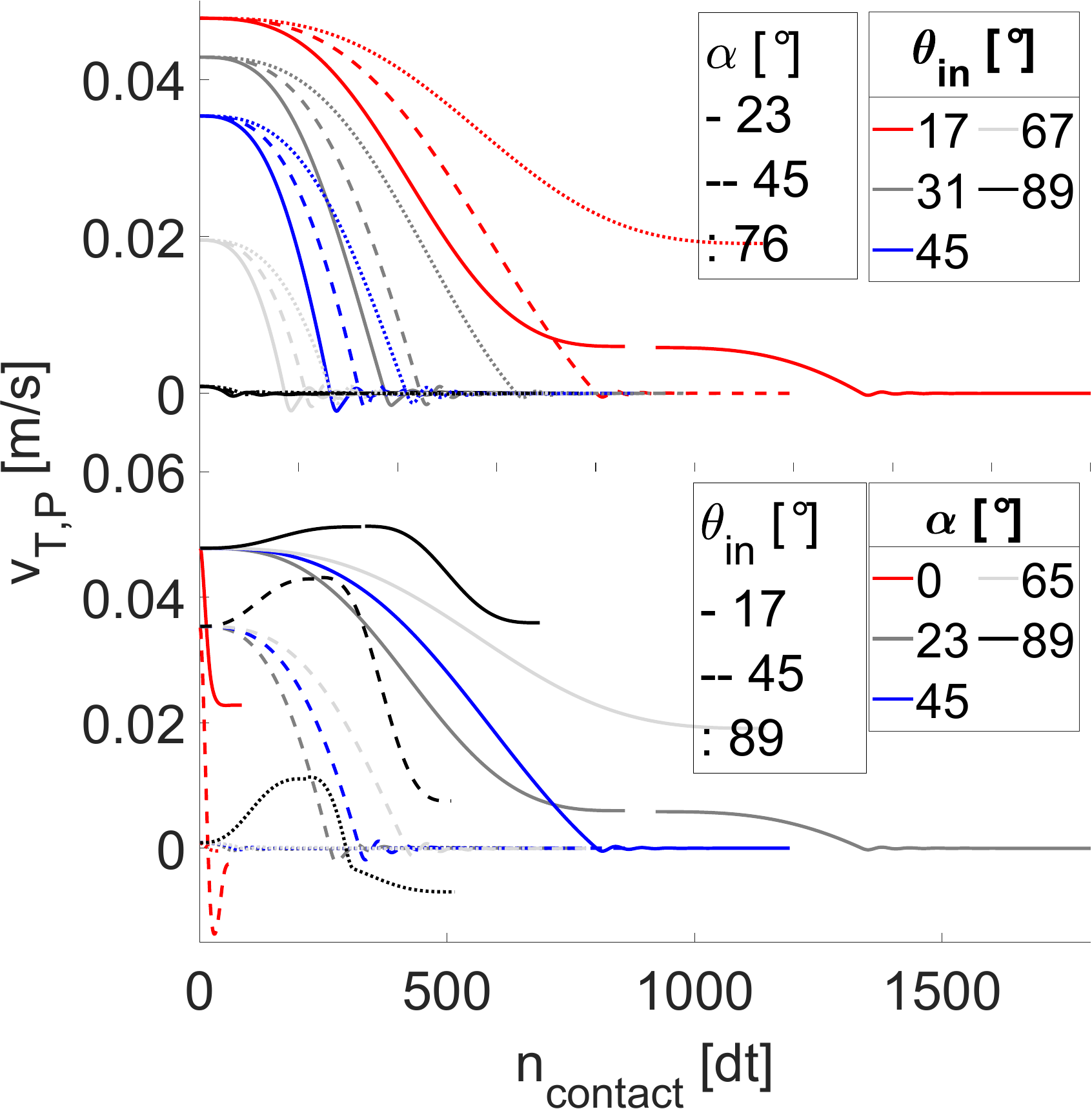}
\caption{Sliding velocity of the contact point, $v_{T,\mathrm{P}}$, for different combinations of orientation $\alpha$ and impact angle $\theta_{\mathrm{in}}$.}
\label{fig:N004_cp_velocity}
\end{figure}

As the collision is initiated without initial rotation, the initial sliding velocity of the particle is only a function of the impact angle, 
\begin{equation}
 v_{T,\mathrm{P}}=v(0)cos{\theta_{\mathrm{in}}}.
\end{equation}
It is larger if the impact is flatter ($\theta_{\mathrm{in}}\rightarrow0^{\circ}$), and smaller if the impact is steeper ($\theta_{\mathrm{in}}\rightarrow90^{\circ}$), see Fig.\,\ref{fig:N004_cp_velocity}, top. During contact the sliding velocity declines to a steady value. For most orientations and impact angles, $v_{T,\mathrm{P}}$ converges to zero and the particle will enter a stick-slip phase before separation. For smaller $\theta_{\mathrm{in}}$ the steady value before separation may be finite. If a second contact phase happens, then the sliding velocity will decrease to zero in the second phase, and the particle will enter stick-slip then. In the case of strongly backwards facing particles, $\alpha\sim89^{\circ}$, the sliding velocity may temporarily increase, before it decreases. For strongly backward facing particles and steep impacts $v_{T,\mathrm{P}}< 0$ at detachment is also possible, see Fig.\,\ref{fig:N004_cp_velocity}, bottom. In general, however, most combinations of $\theta_{\mathrm{in}}$ and $\alpha$ will lead to stick-slip phase before separation.

\begin{figure}[t!]
 \centering
 \includegraphics[width=0.95\columnwidth]{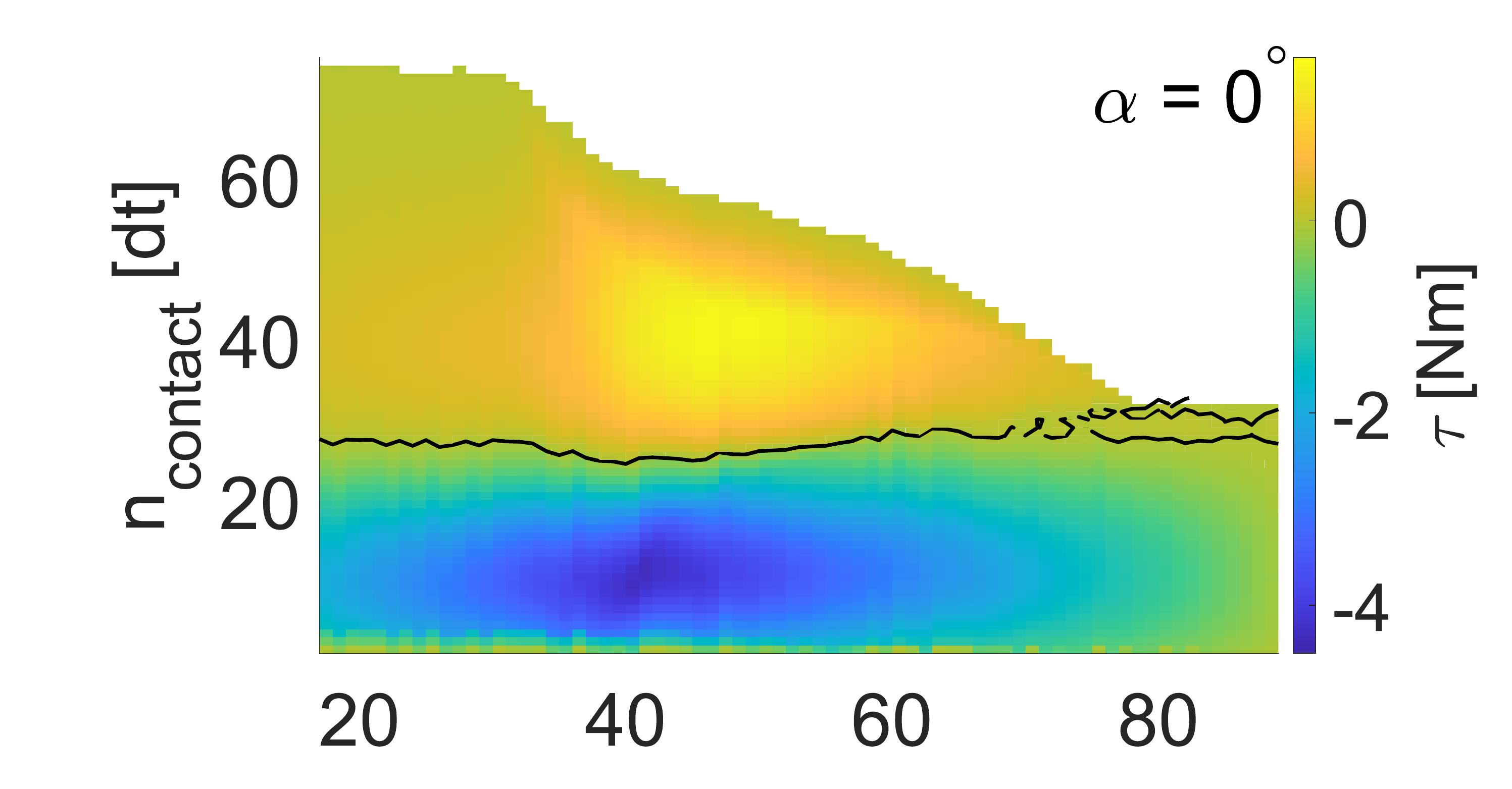}

 \includegraphics[width=0.95\columnwidth]{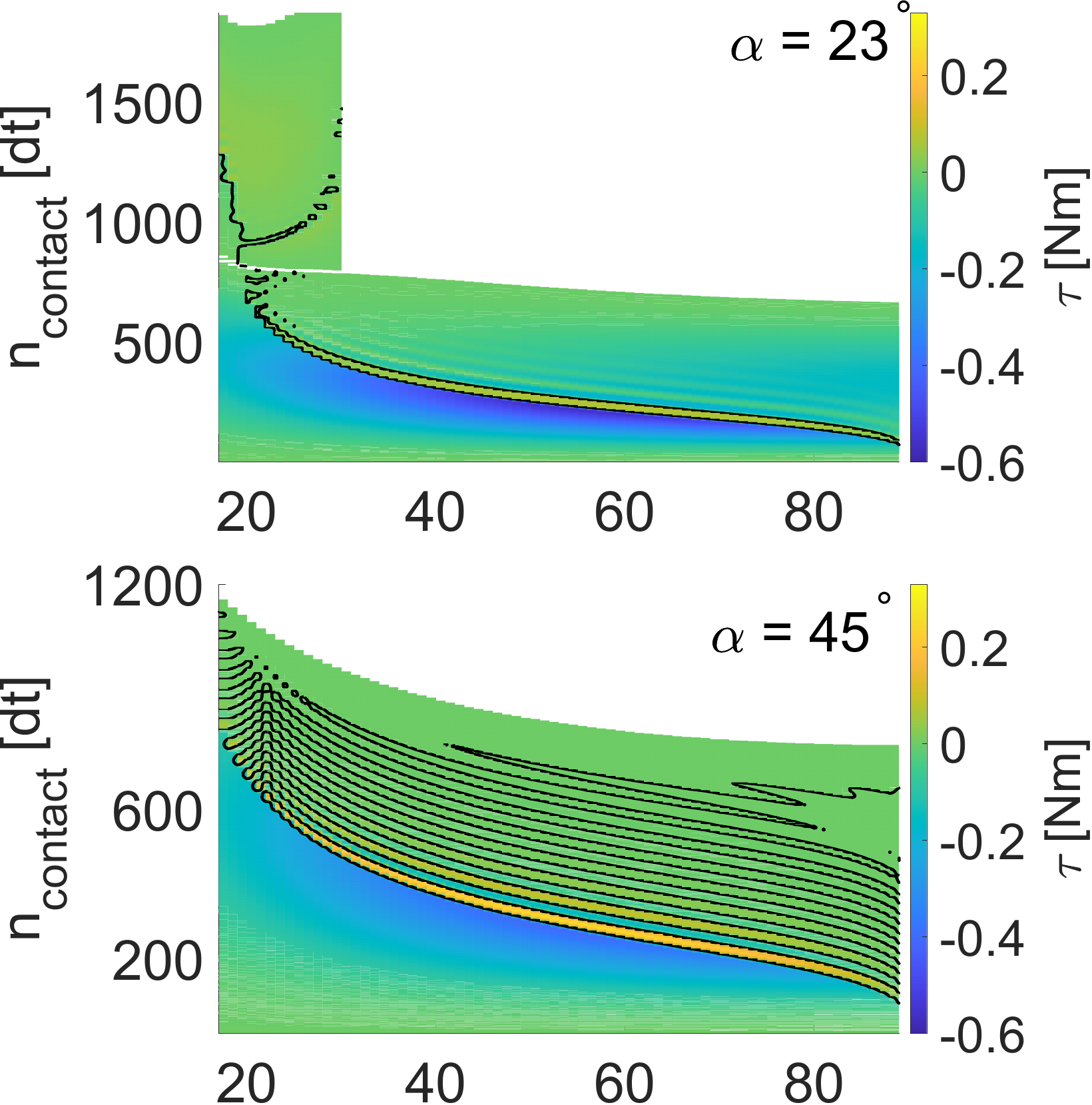}

 \includegraphics[width=0.95\columnwidth]{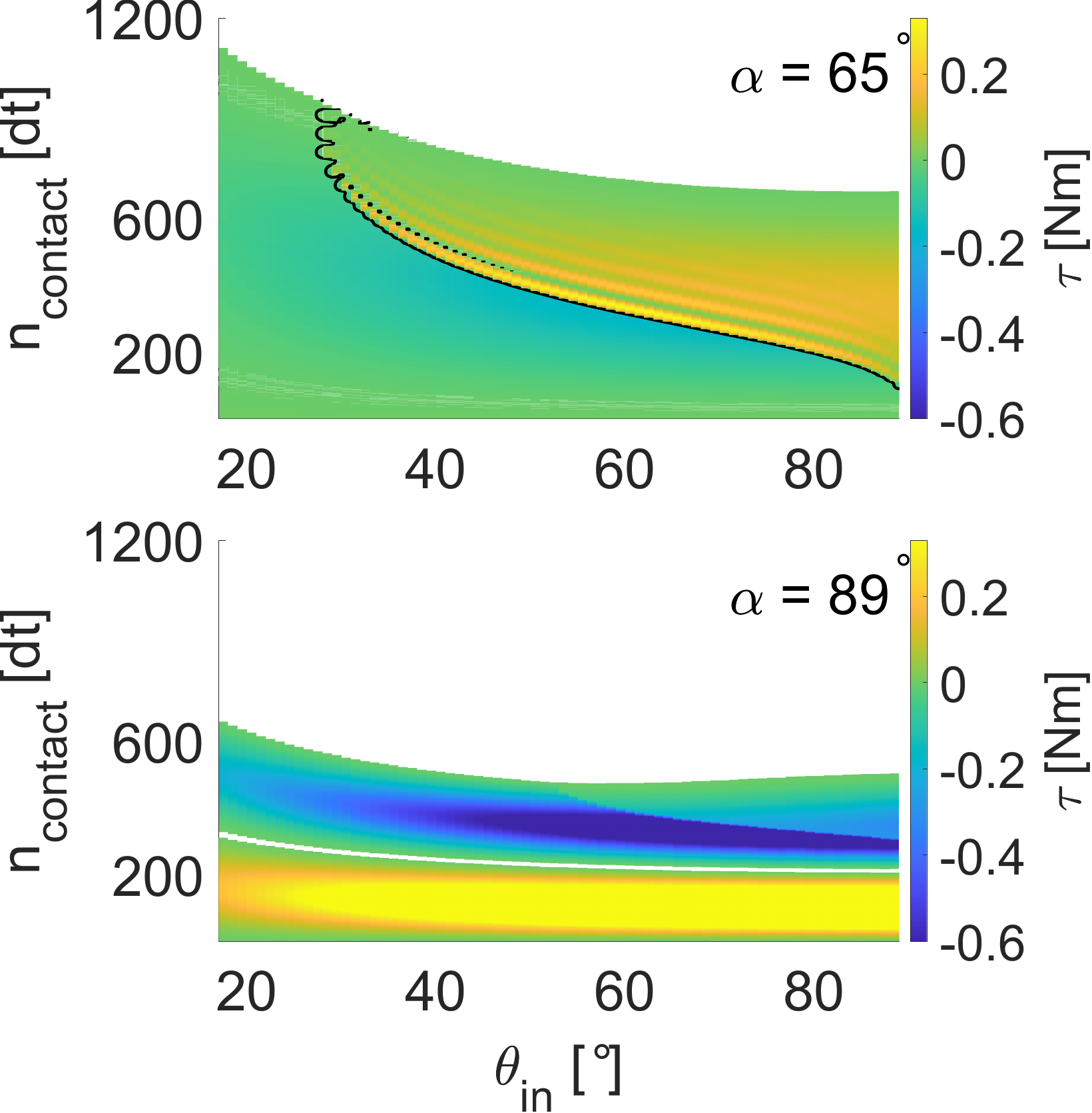}

 \caption{Evolution of the torque $\tau$ for ($\gamma=0.5,\mu=0.5$) during contact and selected initial block orientations. Note that the colour-scale differ and contact duration differ for the different figures. The black line separates positive and negative torque.}
 \label{fig:N004_contact_T}
\end{figure}
As with the disc before, the evolution of the induced torque $\tau$ during contact correlates with the induced tangential force $F_{T}$. In the case of parallel particles ($\theta_{\mathrm{in}} = 0^{\circ}$, a net-torque $\tau=|\bm{r}\times\bm{F}|$ is induced as for oblique impacts the force vector $\bm{F}$ is not parallel to the contact vector $\bm{r}$. For different particle orientation we again observe oscillating patterns of torque, which are predominantly forward rotating ($\tau < 0$) for forward facing particles, backward rotating ($\tau > 0$) for backward facing particles, and fluctuation around zero for $\alpha=45^{\circ}$. For all particle orientations, the particle initially experiences a forward torque ($\tau<0$), the strength of which decreases as $\alpha$ increases. As with the force, for $\alpha=89^{\circ}$, the impact process is a ``double-contact'' again. However, as opposed to forward facing particle orientation, the first impact here induces purely backward rotation, while the second impact induces only forward rotation.

We can write the centre-of-mass velocity of the particle as the sum of its sliding velocity and a contribution of rotation around the contact point,
\begin{equation}
 \bm{v} = \bm{v}_{P} + \bm{\omega}\times\bm{r}_{\mathrm{A}},
\end{equation}
where, due to the choice of the coordinate system, the contact vector $\bm{r}_{\mathrm{A}}$ has a negative component in the y-component and the sign of the x-components depends on the facing of the particle. For a backward facing particle the contact vector is thus
\begin{equation}
 \bm{r}_{\mathrm{A}} = \left(\begin{array}{c}
   +|r_{x}|\\ -|r_{y}|
	 \end{array} \right).
\end{equation}
As in the high $\alpha$, high $\theta_{\mathrm{in}}$ range the particle is rotating backward, $\omega>0$. As shown in Fig.\,\ref{fig:N004_cp_velocity} $\bm{v}_{T,P}>0$ in the this parameter region. Then, the tangential centre-of-mass velocity is
\begin{equation}
 v_{T} = +|v_{T,P}| +(+|\omega|)(-|r_{y}|),
\end{equation}
which is negative if $|v_{T,P}| < (|\omega||r_{y}|)$. As the initial value of $v_{T,P}$ depends on the impact angle $\theta_{\mathrm{in}}$, it is smaller for steeper impacts and larger for flatter impacts, and so more easily reversed if the impact angle is steep, i.e. leading to backwards deflection at higher impact angles. Further, as the decrease of $v_{T,P}$ during contact depends on the response of the tangential force to the impact, it depends primarily on the frictional properties of the material. At the same time, the duration of the impact is proportional to the normal force and changes with the contact damping, where higher damping will lead to longer impacts and thus allows more time for the frictional response in the tangential direction. Therefore, the actual range of $\alpha$ and $\theta_{\mathrm{in}}$ necessary for backward deflection depends on the dissipative parameters of the particle, but always originates from large $\alpha$ and large $\theta_{\mathrm{in}}$.
 
If $|v_{T,P}| > (|\omega||r_{y}|)$, then the particle retains its forward momentum, but still may rotate backwards, as shown previously in Figs.\,\ref{fig:N004_reboundangle},d) and\,\ref{fig:N004_omega},b) for ($\alpha\gtrsim70^{\circ}$ and $\theta_{\mathrm{in}}\lesssim45^{\circ}$).

In contrast, if the particle is facing forward, then
$\bm{r}_{\mathrm{A}} = \left(-|r_{x}|, -|r_{y}|\right)^{\top}$, the particle rotates forward ($\omega<0$) and so the centre-of-mass velocity remains positive.

For discs, the same mechanism is not possible, as the distance from the centre of the disc to its boundary is constant regardless of the orientation and thus rotation cannot induce a new contact.

Lastly, we note that the contact duration significantly varies with the particle orientation and impact angle, even if we omit the ``double-impact'' for forward facing particles. Still, as with discs, impact duration generally decreases as the impact becomes steeper.

\section{Conclusion}
In this work we have systematically studied the relation between particle orientation and impact angle on the planar impact of a disc and a square particle by means of a two-dimensional discrete element method. In particular, we not only focused on the post-impact kinematics of the particle, but also studied the change of the kinematic variables during the impact event itself.
Our investigation was limited to particles without any initial angular velocity to reduce the parameter space. Our overall findings still are consistent with simple hands-on experiences conducted in a previous work and with non-systematic experimental studies in the field of rockfall made by other groups. Below we summarize the main findings of our research. 

The rebound angle of a disc is slightly larger than its impact angle, except for very flat and very steep impacts, but will deviate significantly from equality if rotation of the particle is suppressed. The plot of the induced angular velocity collapses onto the same curve at specific, $\mu$-dependent values of $\theta_{\mathrm{in}}$. Mobilized friction is generally maximal upon impact, but may decrease and even reverse during the contact depending on the impact angle and applied friction coefficient. Similar oscillatoric behaviour during the impact also occurs in the induced torque.

For the planar impact of the block we found a region of backward deflection for a backward facing block at steep impact angles in impacts with dissipative forces and rotational degrees of freedom, and a second zone of large deflection angles for forward facing blocks at primarily low impact angles. We found that these zones are associated with large backward rotation, respectively vanishing angular velocity but large changes to the orientation of the particle. We also note that the kinematics of a disc without rotational degrees of freedom do not equal the kinematics of a square particle with rotation, but only a square particle without rotation.

We further showed that high deflection angles are achieved with a ``double-impact''-kind of event, where the particle detaches from and reconnects with the plane during the collision due to angular momentum obtained upon impact. However, backward deflection follows a different mechanism. For non-parallel particle orientation during impact, the tangential force response and the induced torque show oscillating patters, just as for the disc. In these oscillations, the force remains strictly positive for $\alpha<45^{\circ}$ and strictly negative for $\alpha>45^{\circ}$, and crosses $\alpha=0^{\circ}$ only for $\alpha\sim45^{\circ}$. We consider the consistently negative tangential force together with small initial tangential velocity at large impact angles as the underlying reason for the backward deflection.

The results in this study have been obtained under simplified conditions for impacts without initial angular velocity. As shown in many previous studies and demonstrated by the ``double-impact'' at low impact angles, rotation plays a significant role for the impact kinematics, and its correlation with the particle orientation upon contact and the impact angle needs to be addressed in a future study. Further, collision occur in a three-dimensional environment, which opens up many more possibilities for momentum transfer during and deflection after the impact, which will also be considered in a follow-up work. Another possibility for future work is to extend the framework of this study to the impact kinematics of adhesive particles\,\cite{Thornton1991,Chen2023}.

As this study demonstrated, the rebound kinematics of a particle impact may vary even in a small range of impact conditions, such as particle orientation or impact angle, in a rather complex way. While it is possible to describe successive impacts of discs with the same assumptions, for complex particle shapes each impact ought to be treated individually.

\section{Acknowledgments}
D. Krengel is grateful for the financial support by Professor Takashi Matsushima of the University of Tsukuba and for many helpful comments during earlier stages of this research. D. Krengel is further grateful to Dr. J. Chen of the Japan Agency for Marine-Earth Science and Technology and Professor Hans-Georg Matuttis of The University of Electro-Communications for proofreading the manuscript and many helpful comments.

\section*{Data availability}
Data is available from the corresponding author upon reasonable request.

\appendix
\section{Disc impact}
\subsection{Further coefficients of restitution}

\begin{figure}[h]
 \centering
 \includegraphics[width=\columnwidth]{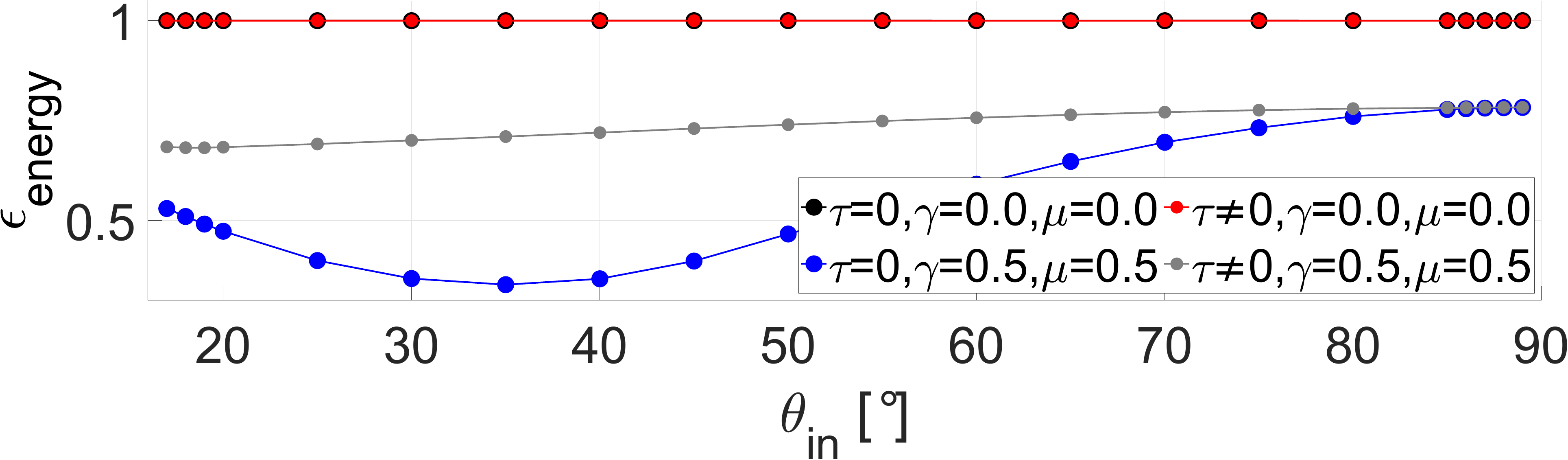}%
 \caption{Energy restitution coefficient $\varepsilon_{\mathrm{energy}}$ as function of the impact angle $\theta_{\mathrm{in}}$ for different conditions.}
 \label{fig:N120_restitution_energy}
\end{figure}

The energy coefficient of restitution $\varepsilon_{\mathrm{energy}}$ (eq.\,\ref{eq:COR_energy}) is largely equivalent to $\varepsilon_{\mathrm{kinematic}}$ and only differs in that for particles with disabled rotation (indicated by $\tau=0$), the minimum is lower than for $\varepsilon_{\mathrm{kinematic}}$, as the velocity contribution is squared, see Fig.\,\ref{fig:N120_restitution_energy}.

\begin{figure}[h]
 \centering
 \includegraphics[width=\columnwidth]{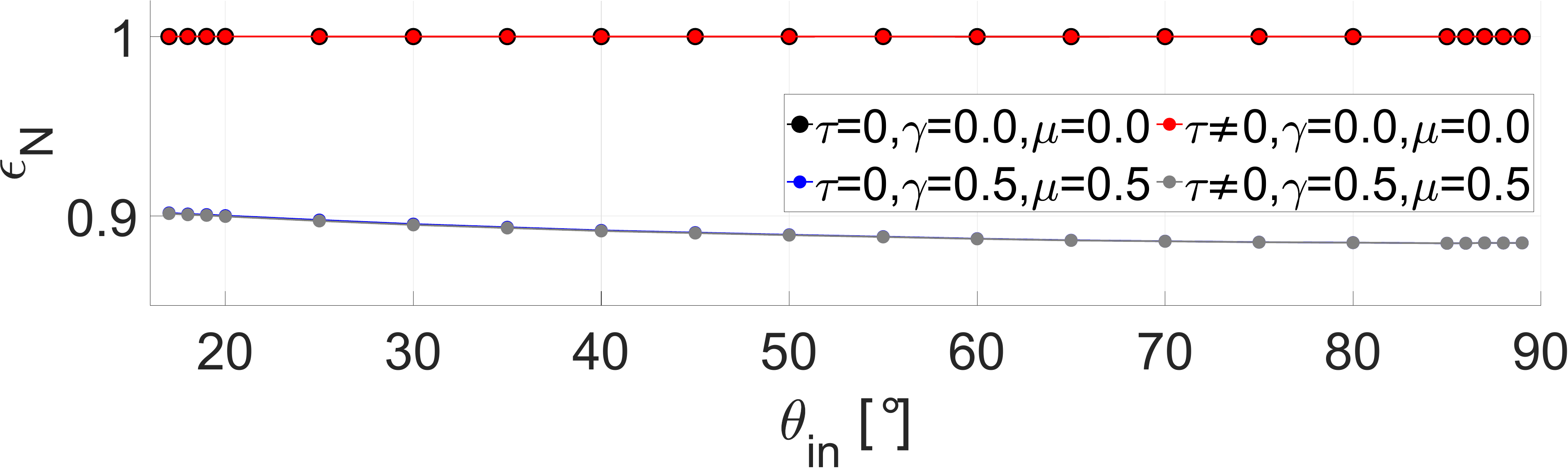}%
 \caption{Normal restitution coefficient $\varepsilon_{\mathrm{N}}$ as function of the impact angle $\theta_{\mathrm{in}}$ for different conditions.}
 \label{fig:N120_restitution_normal}
\end{figure}

The normal coefficient of restitution $\varepsilon_{N}$ (eq.\,\ref{eq:COR_normal}) shows a slight decrease with increasing $\theta_{\mathrm{in}}$ as the normal component of the impact velocity becomes more prominent, which also agrees to some extend with impacts of spheres with initial angular velocity obtained by Asteriou\,\cite{Asteriou2018}. Otherwise $\varepsilon_{N}$ primarily depends on the presence or absence of dissipative normal forces (Fig.\,\ref{fig:N120_restitution_normal}).

The tangential coefficient of restitution $\varepsilon_{T}$ (eq.\,\ref{eq:COR_tangential}),  for non-frictional particles remains trivially constant at 1. For frictional particles with rotational degrees of freedom $\varepsilon_{T}$ slightly decreases from $\theta_{\mathrm{in}}=17^{\circ}$ to $\theta_{\mathrm{in}}=20^{\circ}$, but remains largely constant with any further increase of $\theta_{\mathrm{in}}$. Interestingly, in \,\cite{Asteriou2018}, Fig. 7c), $\varepsilon_{T}$ as a function of tangential impact velocity appears to show a similar trend for particles with initial angular velocity. For dissipative particles without rotation, $\varepsilon_{T}$ quickly decreases to negative values at around $\theta_{\mathrm{in}}\sim50^{\circ}$, increases again to positive values at around $\theta_{\mathrm{in}}\sim70^{\circ}$, before again decreasing to negative values. This variation in $\varepsilon_{T}$ corresponds to the backwards deflection shown in Fig.\,\ref{fig:N120_reboundangle}, and highlights again the necessity of angular velocity to maintain forward momentum at steeper impact angles.
\begin{figure}[t]
 \centering
 \includegraphics[width=\columnwidth]{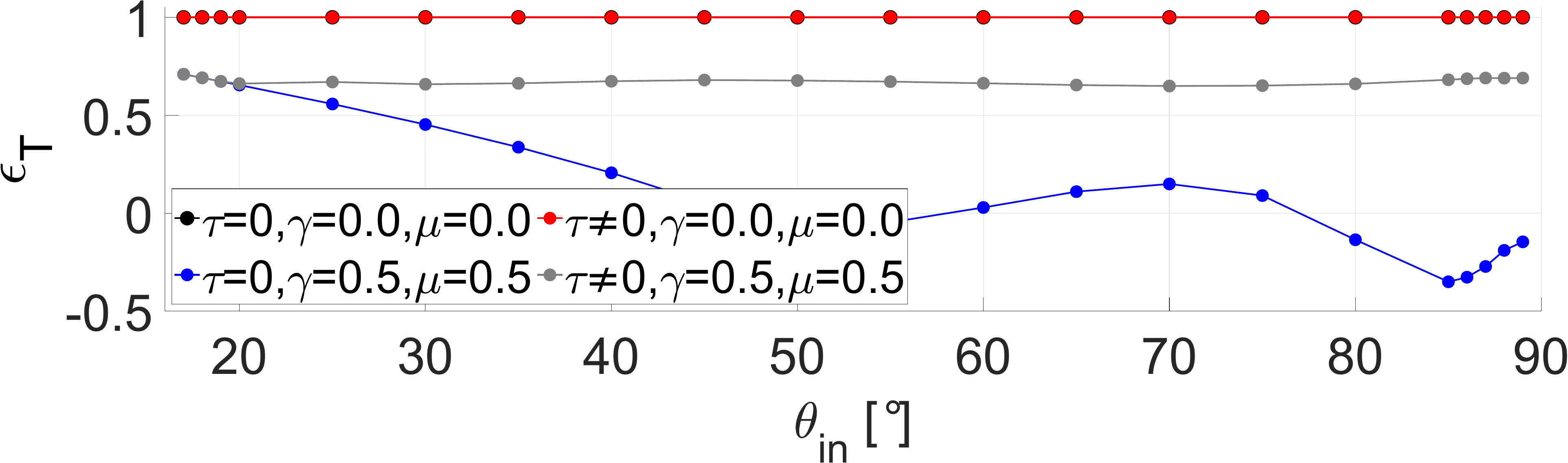}%
 \caption{Tangential restitution coefficient $\varepsilon_{\mathrm{T}}$ as function of the impact angle $\theta_{\mathrm{in}}$ for different conditions.}
 \label{fig:N120_restitution_tangential}
\end{figure}

\subsection{Evolution of the tangential force without rotation}
\begin{figure}[h]
\includegraphics[width=\columnwidth]{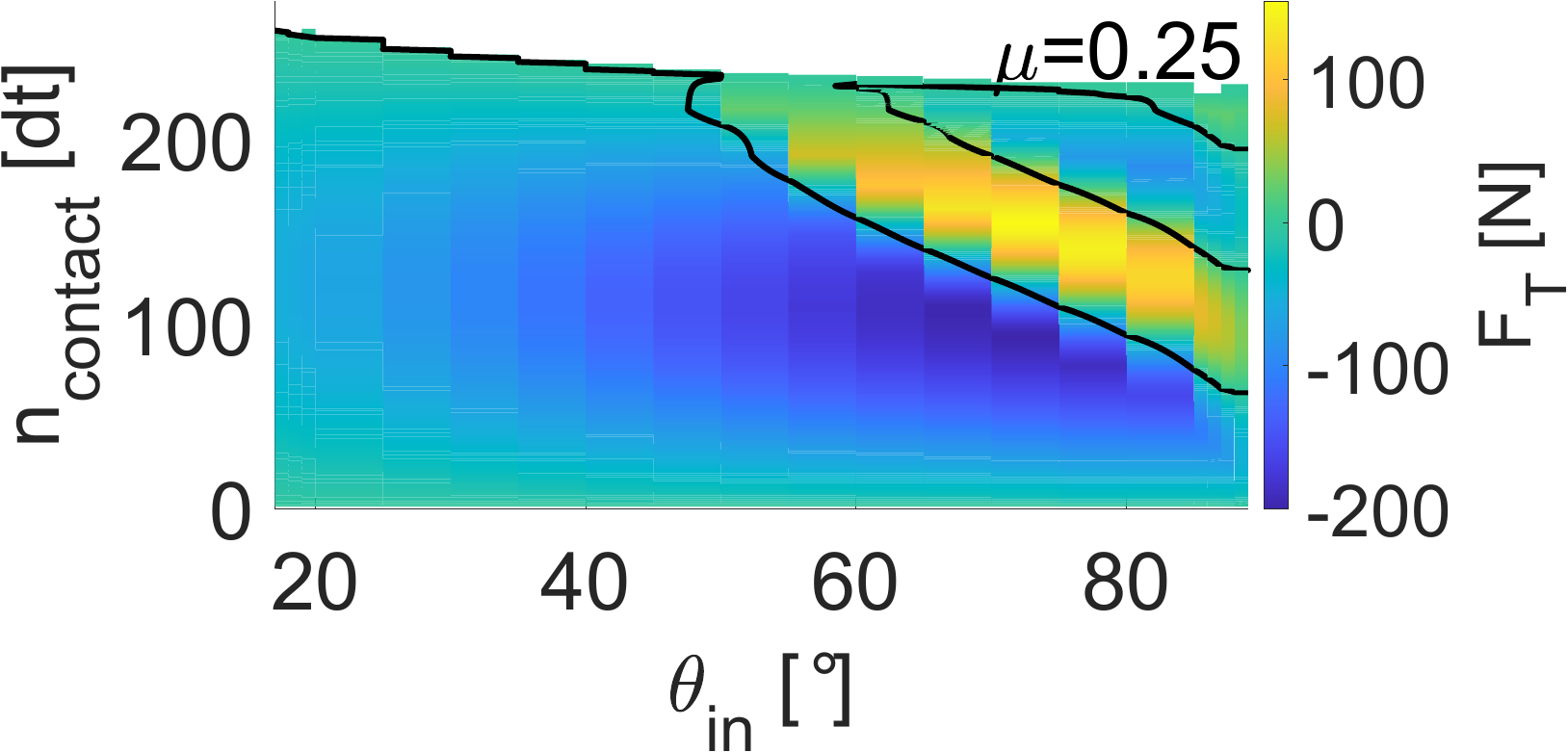}
\caption{Evolution of the tangential force $F_{T}$ acting on the disc during impact with the contact step $n_\mathrm{t}$ for different impact angles $\theta_{\mathrm{in}}$ for a particle with disabled rotation.}
\label{fig:N120_contact_FT_norot}
\end{figure}

In Fig.\,\ref{fig:N120_contact_FT} we showed oscillations in the tangential contact force, depending on impact angle $\theta_{\mathrm{in}}$ and friction coefficient $\mu$. A similar oscillation pattern can be found in dissipative impacts with disabled particle rotation, see Fig.\,\ref{fig:N120_contact_FT_norot}. In this case, the force oscillation starts at higher impact angles and later during the impact process, so that less oscillations occur before separation. The maximum force response, negative and positive, is stronger than in the case with rotation, as the momentum has to be compensated by the rectilinear force and can not be transformed into rotation. 

\section{Square impact}
\subsection{Rebound angle}
For square particles without rotational degrees of freedom, the rebound angle is equivalent to the disc case shown in Fig.\,\ref{fig:N120_reboundangle}. In the absence of any dissipative force, $\theta_{\mathrm{out}}=\theta_{\mathrm{in}}$, trivially (Fig.\,\ref{fig:N004_reboundangle_nodiss},a). The initial orientation of the particle has no influence on $\theta_{\mathrm{out}}$. If dissipative forces are allowed, $\theta_{\mathrm{out}}$ will rapidly approach $90^{\circ}$ and then oscillate back and forth around $\theta_{\mathrm{out}} = 90^{\circ} \pm \delta$, see Fig.\,\ref{fig:N004_reboundangle_nodiss},b). $\theta_{\mathrm{out}}$ is symmetric around $\alpha = 45^{\circ}$ since the overlap area varies with the deviation from $\alpha = 45^{\circ}$, but is constant throughout the contact. Overall, since the impact of a square is sensitive to changes in the orientation, suppressing the ability to rotate can thus reduce the rebound behaviour to that of a disc under equivalent conditions.
\begin{figure}[h]
\includegraphics[width=\columnwidth]{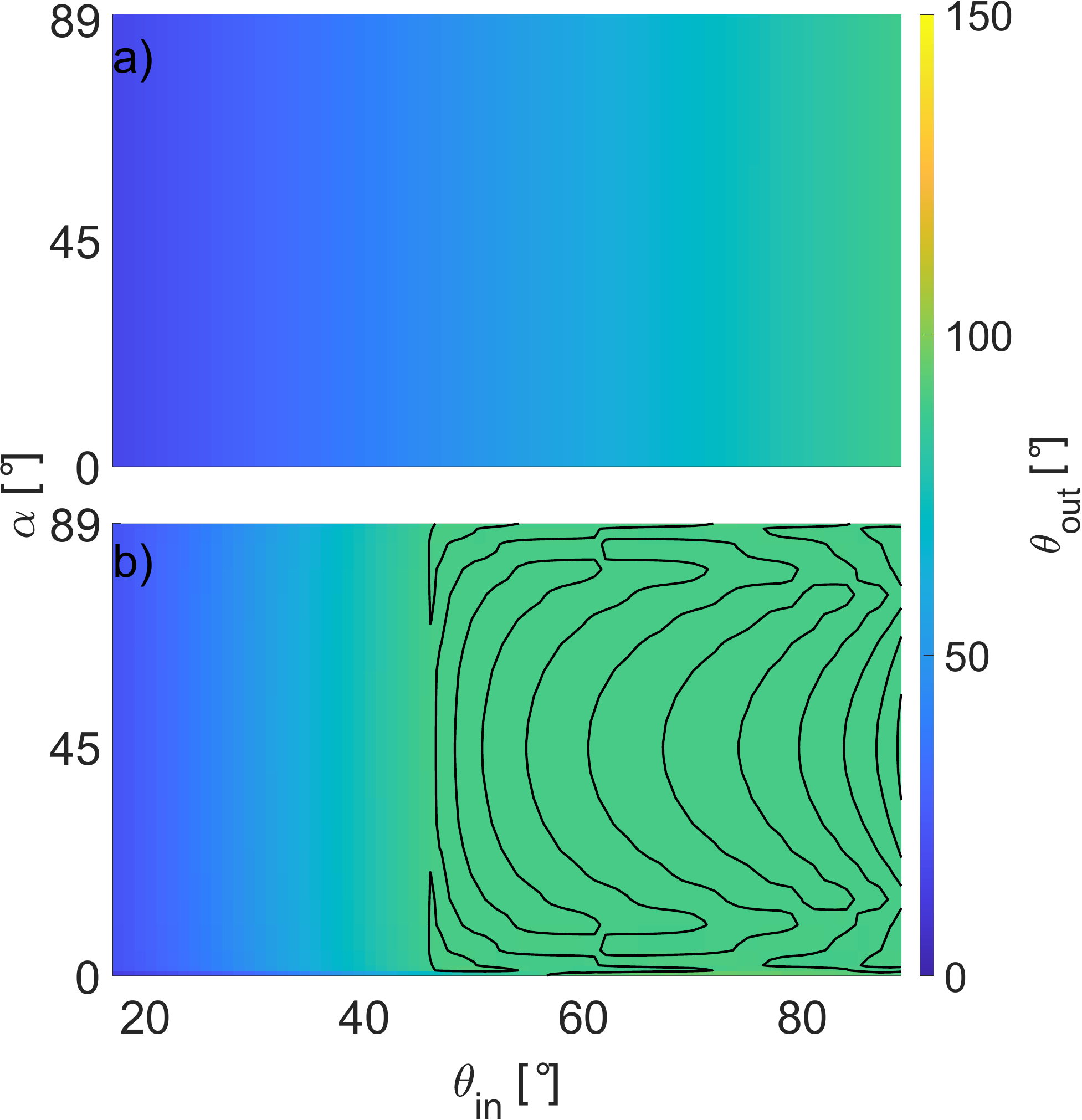}
\caption{Dependence of the rebound angle $\theta_{\mathrm{out}}$ on the impact angle $\theta_{\mathrm{in}}$ and the initial particle orientation $\alpha$ for a particle with inhibited rotation, where a) $\tau = 0, \gamma=0.0,\mu=0.0$, b) $\tau = 0, \gamma=0.5,\mu=0.5$. The black lines denote areas of $\theta_{\mathrm{out}}>90^{\circ}$, i.e. backward deflection.}
\label{fig:N004_reboundangle_norot}
\end{figure} 

\subsection{Coefficients of restitution under different impact conditions}
\begin{figure}[h]
\includegraphics[width=\columnwidth]{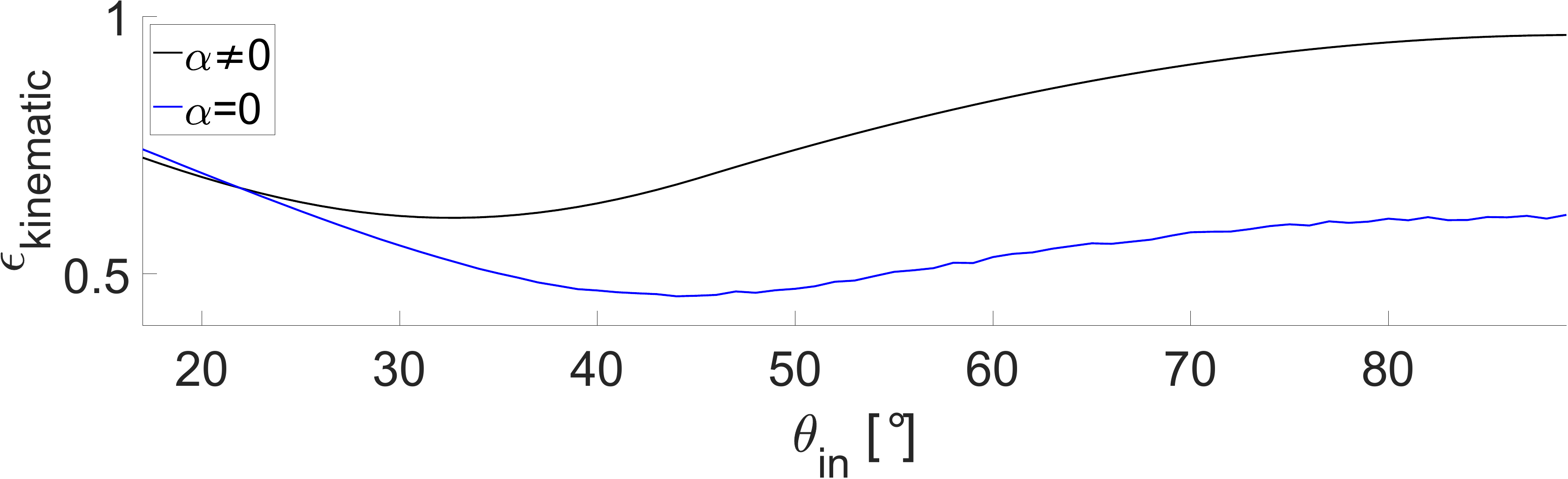}
\caption{Kinematic coefficient of restitution $\varepsilon_{\mathrm{kinematic}}$ for a dissipative particle without rotational degrees of freedom for a parallel impact ($\alpha = 0$) and non-parallel impacts ($\alpha\neq 0$).}
\label{fig:N004_cor_kinematic_norot}
\end{figure}

For a square with inhibited rotation, $\varepsilon_{\mathrm{kinematic}}$ decreases with the impact angle $\theta_{\mathrm{in}}$ to a minimum, then increases again as $\theta_{\mathrm{in}}\rightarrow 90^{\circ}$ (Fig.\,\ref{fig:N004_cor_kinematic_norot}), largely following the behaviour of discs with inhibited rotation (Fig.\,\ref{fig:N120_restitution_kinematic}). In the case of parallel particle orientation, this minimum is both lower and occurs later than for non-parallel orientation. The saturation value at large $\theta_{\mathrm{in}}$ is lower than the value at small $\theta_{\mathrm{in}}$ for parallel impacts, as the larger interface dissipates more energy through the damping component of the normal force. In contrast, for any non-parallel impact $\varepsilon_{\mathrm{kinematic}}(\theta_{\mathrm{in}} \rightarrow 90^{\circ}) > \varepsilon_{\mathrm{kinematic}}(\theta_{\mathrm{in}} \rightarrow 0^{\circ})$.

\begin{figure}[h]
\includegraphics[width=\columnwidth]{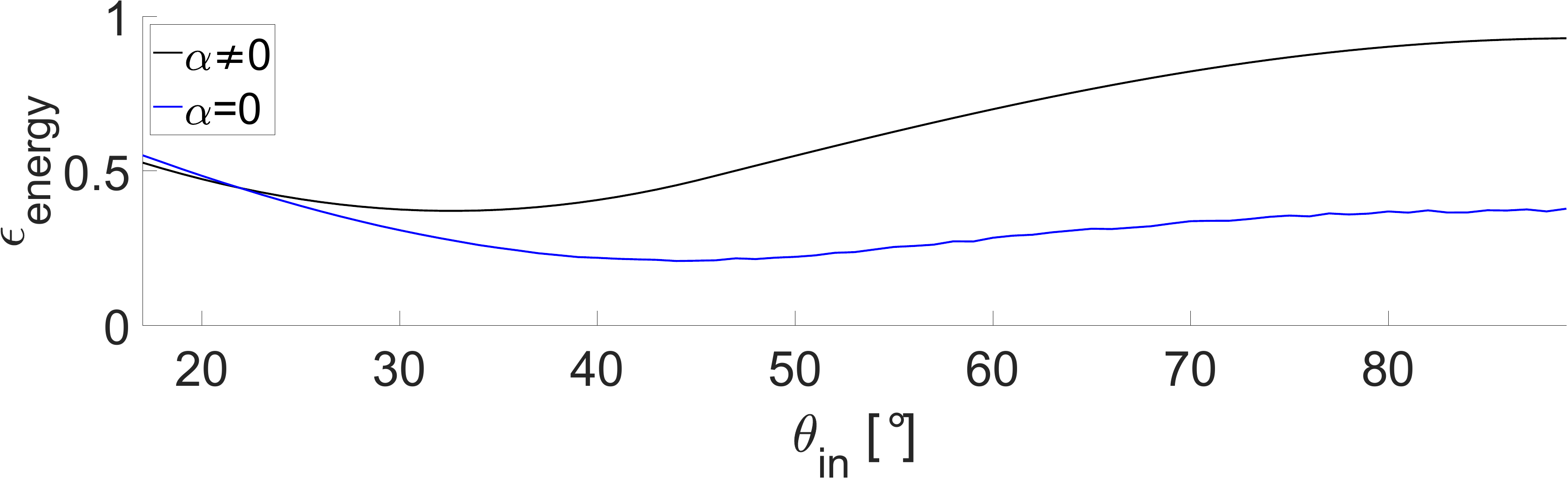}
\caption{Energy coefficient of restitution $\varepsilon_{\mathrm{energy}}$ for a dissipative particle with inhibited rotation for a parallel impact ($\alpha = 0$) and non-parallel impacts ($\alpha\neq 0$).}
\label{fig:N004_COR_energy_norot}
\end{figure}

As with the disc, $\varepsilon_{\mathrm{energy}}$ follows the same functional form as $\varepsilon_{\mathrm{kinematic}}$, where only the magnitude is decreases due to the squared contribution of the velocities (Fig.\,\ref{fig:N004_COR_energy_norot}).

The tangential coefficient of restitution $\varepsilon_{T}$ also follows the functional form for a disc with inhibited rotation (Fig.\,\ref{fig:N120_restitution_tangential}), where an initial steep decrease is followed by oscillations around $\varepsilon_{T} = 0\pm \delta$. However, unlike for discs, these oscillations are symmetric around $\alpha=45^{\circ}$ due to the change in the contact area of the particle. \begin{figure}[h]
\includegraphics[width=\columnwidth]{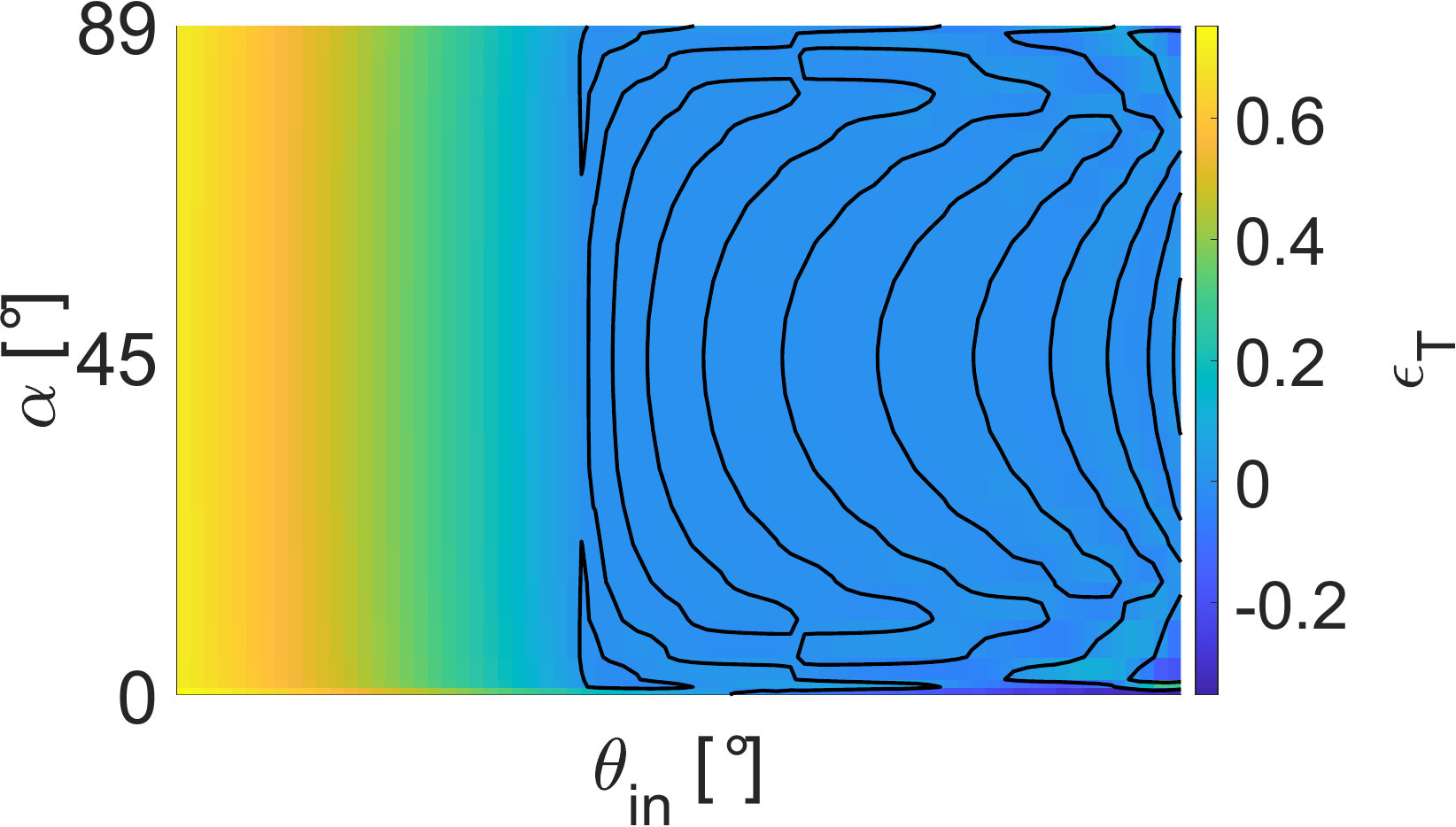}
\caption{Tangential coefficient of restitution $\varepsilon_{\mathrm{T}}$ for a dissipative particle with inhibited rotation for different particle orientation ($\alpha = 0$) and impact angles ($\theta_{\mathrm{in}}$). The black lines mark $\varepsilon_{\mathrm{T}} = 0$}
\label{fig:N004_COR_tangential_norot}
\end{figure}

\bibliography{block_impact}

\end{document}